\documentclass[amsmath,amssymb,12pt,superscriptaddress,nofootinbib]{revtex4-1}
\usepackage{amsmath,amssymb}
\usepackage[pdftex]{graphicx}
\usepackage{color}

\usepackage{ulem}

\renewcommand\b{\begin{equation}}
\newcommand\e{\end{equation}}
 
\newcommand\n{\nonumber}

\newcommand{\pdif}[3]{\frac{{\pd}^{#1} #2}{{\pd} {#3}^{#1}}}

\def\pd{\partial}
\def\dd{{\rm d}}

\def\jT{{\widetilde T}} 
\def\jrho{{\tilde \rho}} 
\def\jp{{\tilde p}} 

\def\av{{\langle \hat\phi_f \rangle_{\rm rms}}}

\def\iz{{\mathbb{Z}}}

\begin{document}
\baselineskip5.5mm
\thispagestyle{empty}

\author{Tomohiro~Nakamura}\email{nakamura.tomohiro@g.mbox.nagoya-u.ac.jp}
\affiliation{
Division of Particle and Astrophysical Science,
Graduate School of Science, Nagoya University, 
Nagoya 464-8602, Japan
}

\author{Taishi~Ikeda}\email{taishi.ikeda@uniroma1.it}
\affiliation{ 
CENTRA, Departamento de F\'{i}sica, Instituto Superior T\'{e}cnico – IST,
Universidade de Lisboa – UL, Avenida Rovisco Pais 1, 1049 Lisboa, Portugal,
 Dipartimento di Fisica} 
\affiliation{
``Sapienza'' Universit\'{a}
di Roma, Piazzale Aldo Moro 5, 00185, Roma, Italy
}

\author{Ryo~Saito}\email{rsaito@yamaguchi-u.ac.jp}
\affiliation{
Graduate School of Science and Engineering, Yamaguchi University, Yamaguchi 753-8512, Japan
}
\affiliation{
Kavli Institute for the Physics and Mathematics of the Universe, Todai Institute for Advanced Study, The University of Tokyo, Chiba 277-8583, Japan (Kavli IPMU, WPI)
}

\author{Norihiro~Tanahashi}\email{tanahashi@phys.chuo-u.ac.jp}
\affiliation{
Department of Physics, Chuo University, Tokyo 112-8551, Japan
}

\author{Chul-Moon~Yoo}\email{yoo@gravity.phys.nagoya-u.ac.jp}
\affiliation{
Division of Particle and Astrophysical Science,
Graduate School of Science, Nagoya University, 
Nagoya 464-8602, Japan
}


\title{Dynamical Analysis of Screening in Scalar-Tensor Theory}

\begin{abstract}

\baselineskip4.5mm
We investigate 
linear and non-linear dynamics of spherically symmetric perturbations on a static configuration 
in scalar-tensor theories focusing on the chameleon screening mechanism. 
We particularly address two questions: how much the perturbations can source the fifth force when the static background is well screened,
and whether the resultant fifth force can change the stability and structure of the background configuration.
For linear perturbations,
we derive 
a lower bound for the square of the Fourier mode frequency $\omega^2$ using the adiabatic approximation.
There may be unstable modes if this lower bound is negative, and we find that 
the condition of the instability 
can be changed by the fifth force although this effect is suppressed by the screening parameter. 
For non-linear perturbations, because we are mainly interested in short wavelength modes
for which the fifth force may become stronger,
we perform numerical simulations 
under the planar approximation. 
For a sufficiently large initial amplitude of the density perturbation, 
we find that the magnitude of the fifth force can be comparable to that of Newtonian gravity 
even when the model parameters are chosen so that the static background is well screened. 
It is also shown that if the screening is effective for the static background, 
the fluid dynamics is mostly governed by the pressure gradient and 
is not significantly affected by the fifth force.

\end{abstract}
\maketitle


\section{Introduction}
\label{intro}

Since the foundation of general relativity (GR), 
alternative theories of gravity have continued to attract much attention, and 
a huge number of theories have been proposed.
In a theoretical point of view, 
those theories are expected to provide some clues for the unified theory of everything. 
In a phenomenological point of view, 
the introduction of such theories is motivated
to explain cosmological riddles 
such as the accelerated expansion of the Universe by the modification.
Due to recent advances in 
observation of gravitational phenomena such as 
gravitational waves (the first signal announced in Refs.~\cite{GW1, GW2}), 
it is expected that we can distinguish a larger class of modified gravity 
theories from the GR in the near future.
Actually, a large class of  modified gravity theories 
has been excluded \cite{GWspeed1,GWspeed2,GWspeed3,GWspeed4} 
by the constraint on the speed of gravitational wave obtained by simultaneous detection of the gravitational wave and the gamma ray signals from a neutron star-neutron star binary \cite{NSGW1,NSGW2}. 
Now, we are 
in the stage to investigate the viability of such theories in more detail.
Detailed theoretical analyses may immediately reject 
modified theories,
or 
bring us predictions on 
new observable quantities that distinguish various theories
which will become available in the near future.

We focus on scalar-tensor theories with an additional scalar field added to GR. 
The additional scalar field may mediate the fifth force through
its coupling to matter fields,  and the existence of the fifth force may contradict 
observations in the solar system or laboratories. 
Then, 
the screening mechanism of the fifth force is necessary for modified gravity theories to evade severe experimental constraints in the solar system 
with allowing large deviations from GR on cosmological scales (see Ref.~\cite{Joyce2015} for review).  
The screening mechanism makes the fifth force small through an environment-dependent parameter: mass, coupling, or kinetic function.
The first class includes the chameleon mechanism \cite{Khoury2004a,Khoury2004}, 
the second class the symmetron mechanism \cite{Hint2010} and environmental dilaton mechanism \cite{Brax2011}, 
and the third class the Vainshtein mechanism \cite{Nicolis2009} and the kinetic screening mechanisms \cite{Babichev2009}. 
In this work, 
we focus on the screening mechanism in the first class: the chameleon mechanism.

The fifth force, or the additional scalar field, can be relevant and should be estimated in any system where the gravity plays an important role.
For $f(R)$ model, in which the chameleon mechanism may work,  numerous efforts to estimate the impact of the fifth force have been made for the large scale structure~\cite{Oyaizu2008f,Oyaizu2008s,Oyaizu2008t,Zhao2011,Zhao2012}, astrophysical objects~\cite{Jain2013,Vikram2014,Terukina2014,Wilcox2016}, the existence of stellar solutions~\cite{Kobayashi2008,Upadhye2009,Tsujikawa:2009yf,Babichev:2009td,Babichev:2009fi}, and so on. 
There are also many attempts to find a screened field in a laboratory by using a specific configuration of a material~\cite{Upadhye2012,Hamilton2015,Lemmel2015,Almasi2015, Burrage2016,Brax2020}. 
In spite of the numerous efforts, 
it has not been fully understood how the chameleon mechanism works in a dynamical system, 
and there are only a few attempts to consider the scalar field dynamics~\cite{Llinares2013, Lindros2016, Hagala2017,Silvestri2011,Upadhye2013}. 
The linear perturbations in scalar-tensor theories 
have been analyzed in many works (e.g.\ \cite{Harada1997,Harada1998}), 
and effects of screening is investigated in Refs.~\cite{Sakstein2013,Brax2017}. 
In Refs.~\cite{Brax2010,Guo2014}, 
the spherical collapse of a matter field 
is calculated for a screened model and $f(R)$ model. 
However, most of these works have been done under the adiabatic approximation, i.e., the amplitude of the scalar field is assumed to trace the minimum of the effective potential and not calculated by solving its dynamical equation.
In the previous works, even when the dynamics of the scalar field is taken into account, 
the matter evolution is given externally and the backreaction due to the fifth force is ignored. 

In this work, we consider the full non-linear dynamics of the fluid and the chameleon field 
in a Newtonian spherically symmetric system. 
To clarify influence of dynamics of the scalar field,
we solve both the matter and scalar field equations simultaneously 
and see how the fifth force is screened in a dynamical system with a spherically symmetric configuration.
In our previous work~\cite{Nakamura2019}, 
we have analyzed the chameleon mechanism in a spherical shell system as an extremely inhomogeneous situation.  
The analysis there suggests that the fifth force could be locally amplified even when the system is screened for the spatially averaged configuration. 
Therefore, 
even when the fifth force is undetectable in the exterior region of a gravitational object, 
the fifth force might be unscreened in the inner region if the object has a highly inhomogeneous configuration. 
Motivated by this fact, 
we investigate how the fifth force is screened in a spherical object with high inhomogeneities and whether the stability and structure of the object are affected.

This work is organized as follows.
In Sec.~\ref{screening}, we briefly review the chameleon mechanism in 
a scalar-tensor theory.
Next, 
in Sec.~\ref{eq_star}, 
we derive basic equations for a system in which a perfect fluid and a scalar field are conformally coupled. 
In Sec.~\ref{linear_an}, 
we analyze the linear stability of a spherical object under the adiabatic approximation but 
taking into account the coupling between the perturbations of the scalar field and fluid,
which has been ignored in the previous works \cite{Sakstein2013}. 
In Sec.~\ref{local_analysis}, we discuss the dynamics of the scalar field in a local region. 
More specifically,
 we first perform a linear analysis and discuss the validity of the adiabatic approximation in Sec.~\ref{linear_adiabatic}. 
Then, in Sec.~\ref{numerical}, 
we perform fully non-linear numerical simulations 
to estimate the magnitude of the fifth force and its backreaction on the structure of the system. 
Finally, we conclude this work in Sec.~\ref{conclusion}, 
discussing the effect of the scalar field on the dynamics of a screened stellar system. 
Throughout this work,
we set 
$c=\hbar=1$ and $M_{\rm pl}^2=1/8\pi G$, where $c$, $\hbar$ and $G$ are the speed of light, reduced Planck's constant and the Newtonian gravitational constant, respectively.

\section{Screening in a scalar-tensor theory}
\label{screening}
Let us briefly review the screening mechanism in a scalar-tensor theory. 
We express the Einstein frame metric by $g_{\mu\nu}$, and the associated Ricci curvature scalar by $R$. 
Then we consider the following action for the gravity, a scalar field $\phi$ and matter fields $\Psi_i$: 
\b
S=\int\dd^4x\sqrt{-g}\left[\frac{M_{\rm pl}^2}{2}R-\frac{1}{2}\nabla_\mu\phi\nabla^\mu\phi-V(\phi)\right]+S_m\left[\Psi_i; 
\tilde g_{\mu\nu}\right],
\label{general_action}
\e
where the potential $V$ is given by an arbitrary function of $\phi$, and 
$\Psi_i$ minimally 
couples with the Jordan frame metric defined as $\tilde{g}_{\mu\nu}\equiv A^2(\phi) g_{\mu\nu}$
with $A(\phi)$ being an arbitrary function of $\phi$. 
From the action~\eqref{general_action},
we obtain the modified Einstein's field equations as 
\b
G_{\mu\nu}=\frac{1}{M_{\rm pl}^2}\left\{\nabla_{\mu}\phi\nabla_{\nu}\phi-g_{\mu\nu}\left(\frac{1}{2}\nabla_{\rho}\phi\nabla^{\rho}\phi+V(\phi)\right)\right\}
+\frac{A^2(\phi)}{M_{\rm pl}^2} \jT_{\mu\nu},
\e
where $\jT_{\mu\nu}$ is the energy-momentum tensor of the matter fields in the Jordan frame defined as
\b
		\jT_{\mu\nu}\equiv-\frac{2}{\sqrt{-\tilde g}}\frac{\delta S_m}{\delta \tilde g^{\mu\nu}} . 
\e
The equation of motion of the scalar field is given by
\b 
\label{eq:scalar field eq}
\Box\phi=V'(\phi)-A'(\phi)A^3(\phi) \jT, 
\e
where $\jT$ is the trace of the energy-momentum tensor of $\jT_{\mu\nu}$ in the 
Jordan frame: $\jT=\tilde{g}^{\mu\nu}\jT_{\mu\nu}$. 
The Bianchi identity leads
\begin{align}
g^{\nu\sigma}\nabla_{\nu} \jT_{\mu\sigma}+2\frac{A'(\phi)}{A(\phi)}g^{\nu\rho}\nabla_{\nu}\phi ~ \jT_{\mu\rho}&=A'(\phi)A(\phi) \jT\nabla_{\mu}\phi. 
\label{bianchi}
\end{align}

As a simple example, 
we take the chameleon model \cite{Khoury2004a,Khoury2004}, where the arbitrary functions are given by
\b
V(\phi)=\frac{\Lambda^{n+4}}{\phi^n},\hspace{1cm}A=e^{\frac{\beta_{\rm c}}{M_{\rm pl}}\phi},
\label{model_chm}
\e 
where $\Lambda$ is a constant with unit mass dimension, $\beta_c$ is a dimensionless constant, and $n$ is an integer.
Moreover, as the background matter, 
we take a non-relativistic fluid with constant density $\rho_c$. 
Then Eq.~(\ref{eq:scalar field eq}) becomes
\b
	\Box\phi=V'(\phi) - \frac{\beta_{\rm c}\rho_c}{M_{\rm pl}} \,,
\e
where we have made the approximation $A\simeq 1+\beta_{\rm c}\phi/M_{\rm pl}$ 
since $\phi/M_{\rm pl}$ is of the order of the gravitational potential or smaller.  
Actually, we suffer from the problem of which frame is physical and the density can be considered as constant, (i.e., the density in the original frame, $g_{\mu\nu}$, is written as $A^{-4}\tilde\rho$) , but
the difference between the frames is irrelevant at this approximation order.
Thus, we have dropped the tilde from the density. 
Then, for this model, we obtain the following effective potential:
\b
		V_{\rm eff}=\frac{\Lambda^{n+4}}{\phi^n} + \frac{\beta_{\rm c}\rho_c }{M_{\rm pl}} \phi \,.
\e
From the effective potential, 
we can also read the minimum point of the potential $\phi_c$ and the mass around the minimum $m_c$ as
\begin{align}
 	\phi_c &\equiv 
\left(\frac{nM_{\rm pl}\Lambda^{n+4}}{\beta_c \rho_c}\right)^{\alpha} \qquad 
\left(\alpha \equiv \frac{1}{n+1}\right)\,,
\label{phimin}
\\
\label{mass}
	m_c^2&\equiv \left. \frac{\dd^2 V_{\rm eff}}{\dd \phi^2} \right|_{\phi=\phi_c} = (n+1)\frac{\beta_c\rho_c}{M_{\rm pl}\phi_c}\propto\rho_c^{\frac{n+2}{n+1}}. 
\end{align}
This dependence on $\rho_c$ guarantees that the scalar field has 
a larger mass and a shorter range of influence in a higher 
 density region. 

Let us review the screening mechanism caused by the $\rho$-dependent mass, i.e.\ the chameleon mechanism, for a spherical star in a static case (see Refs.~\cite{Sakstein2013,Burrage2018review}).
We assume that the density inside the star is much higher than the 
environmental density $\rho_\infty$. 
When the star is static,
we may expect that the scalar field value is trapped at the potential minimum 
well inside the star because the scalar field cannot be excited due to the large effective mass. 
We refer to the typical radius in which the scalar field stays at the potential minimum as screening radius $r_{s}$.
Only the matter outside the screening radius contributes as the source of the scalar field, and then the fifth force on a particle with unit mass outside the star is estimated as
\b
F_\phi\simeq
    2\beta^2\left(1-\frac{\mathcal{M}(r_s)}{\mathcal M(L)}\right)\frac{G\mathcal M(L)}{r^2},
\label{compton}
\e
where $\mathcal{M}(r)\equiv4\pi\int_0^r\dd r' r'^2\rho(r')$ denotes the mass inside the radius $r$. 
Although the fifth force has an amplitude comparable to the Newtonian one for unscreened cases ($r_s\ll L$), it is suppressed by a tiny factor $1-\mathcal{M}(r_s)/\mathcal{M}(L)$  for screened cases ($r_s\sim L$). 
It is known that the screening radius $r_s$ is implicitly estimated from
\b
	\chi\equiv\frac{\phi_{\infty}}{2\beta M_{\rm pl}} \simeq 
		G \int^\infty_{r_s}\dd r \left[ \frac{\mathcal{M}(r)-\mathcal{M}(r_s)}{r^2} \right] \leq |\Phi_N| ,
\label{screen_chi}
\e
where  
$\phi_\infty$ is the potential minimum of the effective potential for a constant density $\rho_\infty$ while $\Phi_N\equiv -G\mathcal{M}(L)/L$ is the Newtonian potential at the surface. 
Note that the parameter $\chi$ is determined by the model parameters as well as the environmental density around the star.
When $\chi$ satisfies $\chi > |\Phi_N|$, 
the condition (\ref{screen_chi}) does not have a solution. 
In this case, 
we cannot use the thin-shell solution (\ref{compton}) and the fifth force becomes comparable to the Newtonian force.
Therefore, 
for screening the fifth force, 
the parameter $\chi$ should satisfy $\chi < |\Phi_N|$
 (see Ref.~\cite{Burrage2018review} for a recent review on experimental bounds).
For the sake of illustration, 
let us estimate the screening radius for a star with a constant density $\rho_c$.
In this case, the screening radius is roughly determined by the Compton wavelength of the scalar field $\lambda_c (\equiv 1/m_c)$ as $r_s^2\simeq L^2 - (\phi_\infty/\phi_c)\lambda_c^2$. 
Hence we may expect that the star is screened if its radius $L$ is much larger than the Compton wavelength $\lambda_c$.

In the previous work \cite{Nakamura2019},  
we have shown that, even if a spherical star is well screened on average, 
the fifth force can be locally comparable to the Newtonian force
when the density is extremely inhomogeneous.
Having this result in mind, in this work we clarify how dynamical density fluctuations affect the dynamics of the scalar field and the fifth force.
When the density becomes dynamical,
the effective potential for the scalar field becomes time-dependent and it will induce nontrivial dynamics of the scalar field and the fifth force. Depending on strength of the time dependence, the screening may be violated and some interesting phenomena associated with the enhanced fifth force could be observed.
Such an enhancement of the fifth force induced by the stellar dynamics is a target of this work.

\section{Basic equations for a spherical system}
\label{eq_star}
In this section, 
we derive basic equations for a non-relativistic spherically symmetric star with keeping the time dependence.
We consider a perfect fluid as the matter field,
\b
	\jT_{\mu\nu}=(\jrho + \jp)\tilde{g}_{\mu\rho}\tilde{g}_{\nu\sigma}\tilde{u}^{\rho}\tilde{u}^{\sigma}+ \jp \tilde{g}_{\mu\nu},
\e
where $\jrho$, $\jp$, and  $\tilde u^{\mu}\equiv A^{-1}(1, \vec{v})$ are the density, pressure, and fluid four velocity in the Jordan frame, respectively.
Hereafter, 
we work in the Newtonian limit, 
i.e.  
$g_{\mu\nu}\dd x^\mu\dd x^\nu=-(1+2\Phi_N)\dd t^2+(1-2\Phi_N)\dd r^2+r^2\dd\Omega^2$ with $\Phi_N\ll1$ and  $\jrho \gg \jp$.  
We also assume that the scalar field is the first order quantity in the post-Newtonian expansion, 
$\phi/M_{\rm pl}\sim (v/c)^2$.
Then, from Eq.~\eqref{bianchi}, 
we obtain
	\begin{align}
		\label{time_bianchi}
		\frac{\partial \jrho}{\partial t}+\vec{\nabla}\left(\jrho\vec{v}\right)
			&=-3\frac{\beta(\phi)}{M_{\rm pl}}\jrho\frac{\partial\phi}{\partial t}, \\
		\label{space_bianchi}
		\jrho\left\{\frac{\partial\vec{v}}{\partial t}+(\vec{v}\cdot\vec{\nabla})\vec v\right\}&=-\vec{\nabla}\jp-
			\jrho\vec{\nabla}\Phi_N-\frac{\beta(\phi)}{M_{\rm pl}}\jrho\vec{\nabla}\phi,
	\end{align}
with $\beta\equiv M_{\rm pl}\dd \ln A(\phi)/\dd\phi$ at the Newtonian order. 
By introducing the ``conserved'' density and pressure as 
$\rho_* \equiv A^{3}\jrho$ and $p_*\equiv A^{3}\jp$, 
 the right-hand side of Eq.~\eqref{time_bianchi} can be absorbed 
into the left-hand side with $\jrho \rightarrow \rho_*$. 
The form of Eq.~\eqref{space_bianchi} remains the same  at the leading order even when we use $\rho_*$ and $p_*$.
As for the gravitational field, 
ignoring the contributions of the scalar field to the energy-momentum tensor as  higher order 
terms in the post-Newtonian expansion, 
we obtain the Poisson's equation,
\b
\nabla^2\Phi_{N}=\frac{1}{2M_{\rm pl}^2}A\rho_*.
\e 
The scalar field equation
in the Newtonian limit is obtained as
\b
\Box\phi=V'(\phi)+\frac{\beta(\phi)}{M_{\rm pl}}A\rho_*.
\e
Hereafter, we assume that the function $A$ can be expanded as $A \simeq 1+\beta\phi/M_{\rm pl}+{\cal O}((\phi/M_{\rm pl})^2)$
with $\beta$ being a constant. 
Then, neglecting the higher-order terms in the post-Newtonian expansion, we have 
the following four basic equations in this paper,
\begin{align}
\label{continuous}
\frac{\partial\rho}{\partial t}+\vec{\nabla}\cdot\left(\rho\vec{v}\right)&=0, \\
\label{euler_equation}
\rho\left\{\frac{\partial\vec{v}}{\partial t}+(\vec{v}\cdot\vec{\nabla})\vec{v}\right\}&=-\vec{\nabla}p-\rho\vec{\nabla}\Phi_{N}-\frac{\beta}{M_{\rm pl}}\rho\vec{\nabla}\phi, \\
\label{poisson}
\nabla^{2}\Phi_N&=\frac{1}{2M_{\rm pl}^2}\rho, \\
\label{eq:scalar}
\Box\phi&=V'(\phi)+\frac{\beta}{M_{\rm pl}}\rho,
\end{align}
where we omit the subscript $*$ for simplicity hereafter. 
Implementing the equation of state, we can determine the full dynamics of the system from these equations. 

Let us consider
a spherically symmetric system, and divide the quantities into the static background values and perturbations as 
\begin{align}
p(t, r)&=p_0(r)+p_1(t, r), \\
\rho(t, r)&=\rho_0(r)+\rho_1(t, r), \\
\Phi_N(t, r)&=\Phi_{N0}(r)+\Phi_{N1}(t, r), \\
\phi(t, r)&=\phi_0(r)+\phi_1(t, r). 
\end{align}
The background variables satisfy the following static equations:
\begin{align}
\label{bg_equil}
\frac{\dd p_0}{\dd r}+\rho_0\frac{\dd\Phi_{N0}}{\dd r}+\frac{\beta}{M_{\rm pl}}\rho_0\frac{\dd\phi_0}{\dd r}=0, \\
\label{bg_poisson}
\frac{1}{r^2}\frac{\dd}{\dd r}\left(r^2\frac{\dd\Phi_{N0}}{\dd r}\right)=\frac{1}{2M_{\rm pl}^2}\rho_0, \\
\label{bg_scalar}
\frac{1}{r^2}\frac{\dd}{\dd r}\left(r^2\frac{\dd\phi_0}{\dd r}\right)=V'(\phi_0)+\frac{\beta}{M_{\rm pl}}\rho_0,  
\end{align} 
while the perturbation variables satisfy the following dynamical equations: 
\begin{align}
\label{per_cont}
\frac{\partial\rho_1}{\partial t}+\frac{1}{r^2}\frac{\partial}{\partial r}\left(r^2(\rho_0+\rho_1)\frac{\partial\delta r}{\partial t}\right)&=0, \\
\label{per_euler}
(\rho_0+\rho_1)
\left(
\frac{\partial^2\delta r}{\partial t^2}+\frac{\partial\delta r}{\partial t}\frac{\partial^2\delta r}{\partial t\partial r}
\right)
&=-\frac{\partial p_1}{\partial r}-(\rho_0+\rho_1)\frac{\partial \Phi_{N1}}{\partial r}-\rho_1\frac{\partial\Phi_{N0}}{\partial r} \n \\
&~~~ 	-\frac{\beta}{M_{\rm pl}} (\rho_0+\rho_1)\frac{\partial \phi_1}{\partial r} - \frac{\beta}{M_{\rm pl}}\rho_1\frac{\partial\phi_0}{\partial r}
	, \\ 
\label{per_poisson}
\frac{1}{r^2}\frac{\partial}{\partial r}\left(r^2\frac{\partial\Phi_{N1}}{\partial r}\right)&=\frac{1}{2M_{\rm pl}^2}\rho_1, \\
\label{per_scalar}
-\frac{\partial^2\phi_1}{\partial t^2}+\frac{1}{r^2}\frac{\partial}{\partial r}\left(r^2\frac{\partial\phi_1}{\partial r}\right)&=V'(\phi_0+\phi_1)-V'(\phi_0)+\frac{\beta}{M_{\rm pl}}\rho_1, 
\end{align}
where we have defined a new variable $\delta r$ as  
\b
\frac{\partial\delta r}{\partial t} \equiv v_r \,
\e
with $v_r$ being the radial component of the velocity. 
It should be noted that the perturbation equations \eqref{per_cont}--\eqref{per_scalar} are exact and the perturbation variables are not necessarily smaller than the background variables.

\section{Linear analysis in the whole spherical region}
\label{linear_an}
First of all, we analyze the background equations assuming the polytropic equation of state
$p=K\rho^\gamma$ and the potential form  $V(\phi)=\Lambda^{n+4}/\phi^n$. The background equations \eqref{bg_equil}--\eqref{bg_scalar} reduce to 
the Lane-Emden-like equation,
\begin{align}
\label{m_lane_emden}
    \frac{1}{\xi^2}\frac{\dd}{\dd\xi}\left(\xi^2\frac{\dd\theta}{\dd\xi}\right)&=-(1+2\beta^2)\theta^l+2\beta^2\frac{1}{\hat\phi_0^{n+1}} \,,
\end{align}
where $\rho_0 = \rho_c\theta^l$ and $\xi\equiv\sqrt{\rho_c^2/(2M_{\rm pl}^2p_c(l+1))}r$ with $\gamma \equiv 1+1/l$, and
\begin{align}
\label{s_lane_emden}
    \frac{1}{\xi^2}\frac{\dd}{\dd\xi}\left(\xi^2\frac{\dd\hat\phi_0}{\dd\xi}\right) &= 
    \frac{2\beta M_{\rm pl}p_c}{\phi_c\rho_c}(l+1)\left(\theta^l - \frac{1}{\hat\phi_0^{n+1}}\right),
\end{align}
where 
$\hat\phi_0 \equiv \phi_0/\phi_c$ with $\phi_c$ given in Eq.~\eqref{phimin}. 
The quantities with a subscript $c$ represent the corresponding quantities at the center of the star. 
The quantity $\theta(\xi)$ parametrizes the radial dependence of the density, and it 
takes $1$ at the center and $0$ at the surface. 
By solving Eqs.~\eqref{m_lane_emden} and \eqref{s_lane_emden}, 
we obtain the surface $\xi=\xi_s$ as the innermost point at which $\theta(\xi)=0$.  
The value of $\xi_s$ is related to the radius of the star $L$ as
\b
    L^2=\frac{M^2_{\rm pl}p_c}{\rho_c^2}2(l+1)\xi_s^2 \,.
\label{prho_L}
\e
Here, let us introduce the screening parameter $\epsilon$ as
\b
    \epsilon \equiv \frac{1}{\beta} \left(\frac{\phi_c}{M_{\rm pl}}\right) \left( \frac{\rho_c L^2}{M_{\rm pl}^2}\right)^{-1} = \frac{1}{2\beta \bar{\Phi}_c} \left(\frac{\phi_c}{M_{\rm pl}}\right) 
\,,
    \label{eq:screening parameter}
\e
where 
$\bar{\Phi}_c \equiv \rho_c L^2/(2M_{\rm pl}^2). $
This screening parameter corresponds to the ratio of the Compton wavelength to the radius of the star, $\epsilon\sim(1/m_cL)^2$  (see also the text below Eq.~(\ref{screen_chi})). 
Using the screening parameter $\epsilon$, the field equation (\ref{s_lane_emden}) is written as 
    \begin{align}
    \frac{1}{\xi^2}\frac{\dd}{\dd\xi}\left(\xi^2\frac{\dd\hat\phi_0}{\dd\xi}\right) &= 
    \frac{1}{\epsilon}\frac{1}{\xi_s^2}\left(\theta^l - \frac{1}{\hat\phi_0^{n+1}}\right).
\end{align}
The star is well screened when $\epsilon \ll 1$. In this case, 
the background scalar field $\hat{\phi}_0$ is approximately fixed at 
the potential minimum as
    \begin{align}
        \hat{\phi}_0 \simeq \hat{\phi}_{\rm min} \equiv \theta^{-\alpha l} \quad \left( \alpha \equiv \frac{1}{n+1} \right) \,,
        \label{phi_min}
    \end{align}
except in the vicinity of the surface.

Linearizing Eqs.~\eqref{per_cont}--\eqref{per_scalar} with respect to 
the perturbation variables, we obtain
\begin{align}
\frac{\partial\rho_1}{\partial t}&=-\frac{1}{r^2}\frac{\partial}{\partial r}\left(r^3\rho_0\frac{\partial\zeta}{\partial t}\right),
\label{cont_eq_linear}
\\
\rho_0r\frac{\partial^2\zeta}{\partial t^2}&=-\frac{\partial p_1}{\partial r}-\rho_1\frac{\dd\Phi_{N0}}{\dd r}-\rho_0\frac{\partial\Phi_{N1}}{\partial r}
-\frac{\beta}{M_{\rm pl}}\left(\rho_0\frac{\partial \phi_1}{\partial r}+\rho_1\frac{\dd \phi_0}{\dd r}\right), 
\label{euler_eq_linear}\\
0&=\frac{1}{r^2}\frac{\partial}{\partial r}\left(r^2\frac{\partial \Phi_{N1}}{\partial r}\right)-\frac{1}{2M^2_{\rm pl}} \rho_1, 
\label{poisson_eq_linear}
\\
\frac{\partial^2\phi_1}{\partial t^2}&=\frac{\partial^2\phi_1}{\partial r^2}+\frac{2}{r}\frac{\partial\phi_1}{\partial r}-\left.\frac{\dd^2V}{\dd\phi^2}\right|_{\phi=\phi_0}\phi_1-\frac{\beta}{M_{\rm pl}}\rho_1,
\label{scalar_eq_linear}
\end{align}
where we have introduced $\zeta\equiv\delta r/r$.
By integrating Eq.~\eqref{cont_eq_linear} with respect to $t$, we find 
\begin{align}
\rho_1&=-\frac{1}{r^2}\frac{\partial}{\partial r}\left(r^3\rho_0\zeta\right),
\label{rho1}
\end{align}
where we have fixed an arbitrary function of $r$ by requiring $\rho_1=0$ for 
$\zeta=\partial\zeta/\partial r=0$. 
Substituting Eq.~\eqref{rho1} into the Poisson equation \eqref{poisson_eq_linear}, 
we obtain
\begin{align}
\frac{\partial}{\partial r}\left(r^2\frac{\partial \Phi_{N1}}{\partial r}+\frac{1}{2M^2_{\rm pl}}\rho_0r^3\zeta\right)&=0.
\end{align}
The regularity at the center $\left.\partial\zeta/\partial r\right|_{r=0}=\left.\partial\Phi_N/\partial r\right|_{r=0}=0$ leads to the following expression:
\b
\frac{\partial \Phi_{N1}}{\partial r}=-\frac{1}{2M_{\rm pl}^2}\rho_0 r\zeta. 
\e
Differentiating the equation of state $p=p(\rho)$, we obtain
\b
\frac{\partial p}{\partial t}+\frac{\partial\delta r}{\partial t}\frac{\partial p}{\partial r}=\frac{\dd p}{\dd \rho}\left(\frac{\partial \rho}{\partial t}+\frac{\partial\delta r}{\partial t}\frac{\partial \rho}{\partial r}\right).
\e
After the integration over $t$, at the first order in the perturbation variables, the above equation becomes 
\begin{align}
p_1&=-p_0\Gamma_0r\frac{\partial\zeta}{\partial r}-\left(\frac{\dd p_0}{\dd r}+p_0\Gamma_0\frac{3}{r}\right)r\zeta,
\end{align}
where $\Gamma_0\equiv\dd p_0/\dd\rho_0\cdot\rho_0/p_0$.
Substituting this equation and the background equations into the Euler equation \eqref{euler_eq_linear} and the scalar field equation \eqref{scalar_eq_linear}, 
we obtain the following two equations: 
\begin{align}
\rho_0\frac{\partial^2\zeta}{\partial t^2}&=\frac{1}{r^4}\frac{\partial}{\partial r}\left(r^4\Gamma_0p_0\frac{\partial\zeta}{\partial r}\right)+\left(\frac{1}{r}\frac{\dd }{\dd r}[(3\Gamma_0-4)p_0]-\frac{2}{r}\frac{\beta}{M_{\rm pl}}\rho_0\frac{\dd \phi_0}{\dd r}-\frac{\beta}{M_{\rm pl}}\rho_0\frac{\dd^2 \phi_0}{\dd r^2}\right)\zeta\n\\
\label{linear_zeta}
	&~~-\frac{\beta}{M_{\rm pl}}\rho_0\frac{1}{r}\frac{\partial \phi_1}{\partial r}\,,\\
\label{linear_phi}
\frac{\partial^2\phi_1}{\partial t^2}&=\frac{\partial^2\phi_1}{\partial r^2}+\frac{2}{r}\frac{\partial\phi_1}{\partial r}-m_0^2\phi_1+\frac{\beta}{M_{\rm pl}}\frac{1}{r^2}\frac{\partial}{\partial r}\left(r^3\rho_0\zeta\right),
\end{align}
where we have defined the effective mass of the scalar field as  
\b
m^2_0\equiv\left.\frac{\dd^2V}{\dd\phi^2}\right|_{\phi=\phi_0} .
\e

In Ref.~\cite{Sakstein2013}, Eq. \eqref{linear_zeta} was analyzed by ignoring the last term, 
which represents the perturbation in the fifth force imposed on the fluid.
We take this term into account by using the adiabatic approximation 
for the scalar field perturbation, in which the perturbation $\phi_1$ is fixed at the potential minimum for the total density, 
i.e., the last two terms in Eq.~(\ref{linear_phi}) balance:
\b
m_0^2\phi_1\sim\frac{\beta}{M_{\rm pl}}\frac{1}{r^2}\frac{\partial}{\partial r}\left(r^3\rho_0\zeta\right) \,.
\label{eq:adiabatic approx}
\e
We will discuss the validity of the adiabatic approximation in Sec.~\ref{linear_adiabatic} by using a simpler setting and proceed here with assuming it. 
With this approximation,
Eq.~(\ref{linear_zeta})
for $\zeta$  becomes
\begin{align}
\rho_0\frac{\partial^2\zeta}{\partial t^2}&=\frac{1}{r^4}\frac{\dd}{\dd r}\left\{r^4\left(\Gamma_0p_0-\frac{\beta^2}{M_{\rm pl}^2}\frac{\rho_0^2}{m_0^2}\right)\frac{\partial\zeta}{\partial r}\right\}\n\\
&~~+\left\{\frac{1}{r}\frac{\dd }{\dd r}[(3\Gamma_0-4)p_0]
-\frac{\rho_0}{M_{\rm pl}^2}\frac{1}{r}\frac{\dd}{\dd r}\left(\frac{\beta^2}{m_0^2r^2}\frac{\dd}{\dd r}(r^3\rho_0)\right)
\right\}\zeta,\label{approzeta}
\end{align}
where we have omitted terms which contain derivatives of the background scalar field for simplicity (see Ref.~\cite{Sakstein2013} for its effect).  
Expanding
$\zeta$ in Fourier modes as
\b
\zeta=\int \dd\omega \tilde{\zeta}e^{i\omega t}\,,
\e
we obtain  
\begin{align}
0&=\frac{1}{r^4}\frac{\dd}{\dd r}\left\{r^4\left(\Gamma_0p_0-\frac{\beta^2}{M_{\rm pl}^2}\frac{\rho_0^2}{m_0^2}\right)\frac{\partial\tilde\zeta}{\partial r}\right\}\n\\
&~~+\left\{\rho_0\omega^2+\frac{1}{r}\frac{\dd }{\dd r}[(3\Gamma_0-4)p_0]-\frac{\rho_0}{M_{\rm pl}^2}\frac{1}{r}\frac{\dd}{\dd r}\left(\frac{\beta^2}{m_0^2r^2}\frac{\dd}{\dd r}(r^3\rho_0)\right)
\right\}\tilde\zeta . \label{approzeta_omega}
\end{align}
Multiplying the both sides of Eq.~\eqref{approzeta_omega} by $r^4\tilde \zeta$ 
and integrating with respect to $r$ from the center to the surface of the star, we obtain
\begin{align}
\omega^2 \int_0^L {\rm d}r ~ r^4\rho_0 \tilde{\zeta}^2&=\int^L_0 \dd r ~ r^4\Biggl\{\Gamma_0p_0\left(1-\frac{\beta^2\rho_0^2L^2}{M_{\rm pl}^2p_0\Gamma_0}\frac{1}{L^2m_0^2}\right)\left(\frac{\dd\tilde\zeta}{\dd r}\right)^2\n\\
&~~~-\left[\frac{1}{r}\frac{\dd }{\dd r}[(3\Gamma_0-4)p_0] - \frac{\beta^2}{M_{\rm pl}^2}\frac{\rho_0}{m_0^2}\left(\frac{3}{r}\frac{\dd \rho_0}{\dd r}+\frac{\dd^2 \rho_0}{\dd r^2}\right)\right]\tilde\zeta^2\Biggr\}, 
\end{align}
where we have dropped the surface term imposing the regularity at the center and 
an appropriate boundary condition at the surface.  
For a well-screened object 
that satisfies $L^2m_0^2\gg1$, the first term in the right-hand side 
is always positive and we obtain the following inequality: 
\begin{align}
\omega^2&\geq-\frac{\int^L_0\dd r ~ r^4\left\{\frac{1}{r}\frac{\dd }{\dd r}[(3\Gamma_0-4)p_0]-\frac{\beta^2}{M_{\rm pl}^2}\frac{\rho_0}{m_0^2}\left(\frac{3}{r}\frac{\dd \rho_0}{\dd r}+\frac{\dd^2 \rho_0}{\dd r^2}\right)\right\}\tilde\zeta^2}{\int^L_0 {\rm d}r ~ r^4\rho_0\tilde\zeta^2}.
\label{omegaineq}
\end{align}
If we ignore the scalar field contribution proportional to $\beta^2$ in Eq.~\eqref{omegaineq}, this inequality implies that unstable modes with $\omega^2<0$ may be present if $\Gamma_0<4/3$ 
since the pressure and the density are decreasing functions of $r$.
This is a familiar result in the standard Newtonian gravity.
It is worthy to note that, because 
the second term in the braces 
can be positive/negative, 
the scalar field contribution extends/reduces the possible parameter region of the 
instability.
Nevertheless,
the ratio of the second term to the first term can be roughly estimated using Eq.~(\ref{prho_L}) as
\b
		\frac{1}{p_0}\cdot \frac{\beta^2}{M_{\rm pl}^2}\frac{\rho_0^2}{m_0^2} \sim \beta^2\frac{1}{m_0^2L^2} \sim \epsilon \,.
\e
Here, the difference between the quantities with the subscripts $c$ and $0$ has been ignored.
This estimation shows that 
the scalar field contribution
is suppressed by the screening parameter, and 
it cannot drastically change the stability.

\section{Full dynamics in a local region}
\label{local_analysis}

In this section, we investigate the full dynamics of the scalar field and the matter. 
In Ref.~\cite{Nakamura2019}, it was shown that the fifth force can be locally comparable to the Newtonian force 
when the matter distribution is given by static concentric shells, which may be regarded as a system with very strong spatial inhomogeneities. 
In a more realistic setup, 
such a configuration would be unstable and dynamically changes. 
Our interest is how the associated fifth force evolves and affects the dynamics of the matter configuration.
Since the fifth force is expected to be more enhanced for larger amplitude of inhomogeneities, 
we will solve the dynamical equations \eqref{per_cont}--\eqref{per_scalar} including the non-linear regime.

In order to make the analysis more amenable, 
we introduce the planner approximation focusing on the region whose interval is much smaller than the curvature scale:
$R\leq r\leq R+a$ for a certain radius $R$ and a width $a\ll R$. 
We introduce a new radial coordinate $x$ in this short range by $x \equiv r-R$.   
Focusing on only short wavelength modes of perturbations fluctuating in this region,
we assume 
\b
\frac{\partial \Pi_0}{\partial r}\ll\frac{\partial \Pi_1}{\partial r},
\label{planar_approx}
\e
where $\Pi$ denotes $\rho$, $p$, $\Phi_{N}$ or $\phi$. Then Eqs.~\eqref{per_cont}--\eqref{per_scalar} can be reduced as follows:  
\begin{align}
\label{cham_cont}
\frac{\partial\rho_1}{\partial t}+\frac{\partial}{\partial x}\left((\rho_0+\rho_1)\frac{\partial\delta x}{\partial t}\right)&=0, \\
\label{cham_euler}
(\rho_0+\rho_1)\left(\frac{\partial^2\delta x}{\partial t^2}+\frac{\partial\delta x}{\partial t}\frac{\partial^2\delta x}{\partial t\partial x}\right)&=-\frac{\partial}{\partial x}p_1-(\rho_0+\rho_1)\frac{\partial\Phi_{N1}}{\partial x}-\frac{\beta}{M_{\rm pl}}(\rho_0+\rho_1)\frac{\partial\phi_1}{\partial x}, \\
\label{cham_poisson}
\frac{\partial^2\Phi_{N1}}{\partial x^2}&=\frac{1}{2M_{\rm pl}^2}\rho_1, \\
\label{cham_phi}
\left(-\frac{\partial ^2}{\partial t^2}+\frac{\partial ^2}{\partial x^2}\right)\phi_1&=V'(\phi_0+\phi_1)-V'(\phi_0)+\frac{\beta}{M_{\rm pl}}\rho_1 \,.
\end{align}
Hereafter we use 
the polytropic equation of state $p=K\rho^{1+1/l}$
and describe the system using
the variables $\theta$, $\hat \phi_0$ defined in Eqs.~\eqref{m_lane_emden} and \eqref{s_lane_emden} for the background variables. 
We obtain 
    \begin{align}
\label{nl_rho_1}
\frac{\partial\hat\rho_1}{\partial \hat{t}}&=-\frac{\partial}{\partial \hat x}\left((\theta^l+\hat\rho_1)\frac{\partial\hat\zeta}{\partial \hat t}\right),\\
\label{nl_euler_1}
    \frac{\partial^2 \hat\zeta}{\partial \hat{t}^2}&=-\frac{\partial}{\partial \hat x}\left[\frac{1}{2}\left(\frac{\partial\hat\zeta}{\partial t}\right)^2+\frac{p_c}{\rho_c}\left\{(l+1)(\theta^l+\hat\rho_1)^{1/l}+b\hat\Phi_{N1}+2\beta^2b\epsilon\hat\phi_1\right\}\right],\\
\label{nl_poisson_1}
    \frac{\partial^2\hat\Phi_{N1}}{\partial\hat{x}^2}&=\frac{a^2}{L^2}\hat\rho_1,\\
\label{nl_phi_1}
\frac{\partial^2\hat\phi_1}{\partial \hat t^2}&=\frac{\partial^2\hat\phi_1}{\partial \hat{x}^2}-\frac{1}{\epsilon}\frac{a^2}{L^2}\left\{\frac{M_{\rm pl}}{\beta\rho_c}\left(V'(\phi_c(\hat\phi_0+\hat\phi_1))-V'(\phi_c\hat\phi_0)\right)+\hat\rho_1\right\},
    \end{align}
where we have normalized the coordinates and the variables as $\hat{t} \equiv t/a, \hat{x} \equiv x/a, \hat\zeta \equiv \delta x/a, \hat\rho_1 \equiv \rho_1/\rho_c, \hat\phi_1 \equiv \phi_1/\phi_c$, $\hat\Phi_{N1}\equiv \Phi_{N1}/\bar\Phi_c\equiv\Phi_{N1}/(\rho_c L^2/2M_{\rm pl}^2)$
and $b \equiv L^2\rho_c^2/(2p_cM_{\rm pl}^2) = (l+1)\xi_s^2$.  
The derivative of $\theta$ is dropped by applying the planar approximation \eqref{planar_approx}. 
Hereafter, we take $b=1$ for simplicity.

\subsection{Linear analysis and validity of the adiabatic approximation}
\label{linear_adiabatic}

First, we analytically study the dynamics at the linear level and show that the deviation from the adiabatic approximation (\ref{eq:adiabatic approx}), i.e.\ the dynamics of the scalar field, can be important on very small scales. 
Expanding up to the linear order of the perturbation variables, 
Eqs.~\eqref{nl_rho_1}--\eqref{nl_phi_1} becomes
\begin{align}
\label{cham_contl}
    \frac{\partial\hat\rho_1}{\partial \hat t}+\theta^{l}\frac{\partial^2\hat\zeta}{\partial \hat t\partial \hat x}&=0\,, \\
\label{cham_eulerl}
    \frac{\partial^2\hat\zeta}{\partial \hat t^2}&=-\frac{p_c}{\rho_c}\left(\gamma\theta^{1-l}\frac{\partial\hat\rho_1}{\partial x}+\frac{\partial\hat\Phi_{N1}}{\partial \hat x}+2\beta^2\epsilon\frac{\partial\hat\phi_1}{\partial \hat x}\right)\,, \\
\label{cham_poissonl}
\frac{\partial^2\hat\Phi_{N1}}{\partial \hat x^2}&=\frac{a^2}{L^2}\hat\rho_1 \,,\\
\label{cham_phil}
    \left(-\frac{\partial ^2}{\partial \hat t^2}+\frac{\partial ^2}{\partial \hat x^2}\right)\hat\phi_1&=\frac{1}{\epsilon}\frac{a^2}{L^2}\left(\hat m^2(\phi_0)\hat\phi_1+\hat\rho_1\right)\,,
\end{align}
where $\hat m^2(\phi_0)\equiv M_{\rm pl}\phi_c/(\beta \rho_c)\dd^2V/\dd\phi^2|_{\phi=\phi_0}$.
These equations are combined to the following two equations by the same procedure 
used for Eqs.~\eqref{linear_zeta} and \eqref{linear_phi}: 
\begin{align}
\label{app_zeta}
\frac{\partial^2\hat\zeta}{\partial \hat t^2}&= c_s^2 \frac{\partial^2\hat\zeta}{\partial \hat x^2} + \frac{c_s^2}{\gamma\theta}\left(\frac{a^2}{L^2}\theta^l\hat\zeta- 2\beta^2\epsilon\frac{\partial\hat\phi_1}{\partial \hat x}\right) \,, \\
\label{app_phi}
\frac{\partial^2\hat\phi_1}{\partial \hat t^2}&=\frac{\partial^2\hat\phi_1}{\partial \hat x^2}-\frac{1}{\epsilon}\frac{a^2}{L^2}\left(\hat m^2\hat\phi_1-\theta^l \frac{\partial\hat\zeta}{\partial \hat x}\right) \,,
\end{align}
where $c_s^2\equiv\gamma p_0/\rho_0=\gamma\theta p_c/\rho_c$ is the sound speed of the perturbation $\zeta$. 
On the right-hand side of Eq.~(\ref{app_zeta}), 
the first term is the pressure gradient term, and the two terms in the parentheses correspond to the Newtonian force and the fifth force, respectively.
The sound speed can be estimated as
    \begin{align}\label{eq:cs}
        c_s^2 = \gamma\theta p_c/\rho_c \sim \frac{G\mathcal{M}(L)}{L} \,,
    \end{align}
which implies that $c_s^2\ll 1$ for a non-relativistic object and $\zeta$ slowly varies. 

In the planar approximation, we define the adiabatic mode as 
    \begin{align}\label{eq:adiabatic approx linear}
       \hat\phi_1^{({\rm ad})} \equiv \frac{1}{\hat{m}^2}\theta^l\frac{\partial \hat\zeta}{\partial \hat x} \,,
    \end{align}
for which the last term in Eq.~(\ref{app_phi}) vanishes.
This corresponds to Eq.~\eqref{eq:adiabatic approx} for a spherically symmetric star. For this mode,
the scalar field is fixed at the minimum of the effective potential, and it slowly varies as $\hat\zeta$ varies:
    \begin{align}
        \frac{\partial^2 \hat\phi_1^{({\rm ad})}}{\partial \hat t^2} 
        \sim c_s^2 \frac{\partial^2 \hat\phi_1^{({\rm ad})}}{\partial \hat x^2} 
        \ll \frac{\partial^2 \hat\phi_1^{({\rm ad})}}{\partial \hat x^2} \,,
    \end{align}
where we have assumed the pressure term is dominant in Eq.~\eqref{app_zeta} compared to the gravity term to maintain the stability of the system. 
When the adiabatic approximation (\ref{eq:adiabatic approx linear}) is accurate, 
the ratio of the fifth force to the gravitational force in Eq.~(\ref{app_zeta}) is estimated as
    \begin{align}
        F_\phi/F_{\rm Newton} = -2\beta^2
        \epsilon \frac{\partial\hat\phi_1^{({\rm ad})}}{\partial \hat x} \bigg/ \frac{a^2}{L^2}\theta^l\hat\zeta \simeq  
        -2\beta^2 \frac{1}{\hat{k}_{\rm ad}^2} \left(\frac{1}{\hat\zeta}\frac{\partial^2 \hat\zeta}{\partial \hat x^2}\right) \,,
    \end{align}
where we have introduced
  \begin{align}
        \hat k_{\rm ad} \equiv \frac{a}{L}\hat m\sqrt{\frac{1}{\epsilon}} = a \sqrt{V''(\phi_0)} \,.
    \end{align}
For a Fourier component with a wavenumber $\hat k$, the ratio is of the order of $\beta^2 (\hat{k}/\hat{k}_{\rm ad})^2$. 
This result indicates that the fifth force is comparable to the gravitational force when the small-scale modes with $\hat k \gg \hat k_{\rm ad}$ are excited. However, as we will see in the following,  we need to take into account the violation of the adiabatic approximation, i.e.\ the dynamics of the scalar field, in this regime.

To see the deviation from the adiabatic mode (\ref{eq:adiabatic approx linear}), it would be convenient to introduce the idea of ``slow" and ``fast" modes. 
Since the sound speed of the density perturbation $c_s$ is small (see Eq.~(\ref{eq:cs})), 
there is a large hierarchy between the variational time scales of two eigenmodes of the simultaneous differential equations (\ref{app_zeta}) and (\ref{app_phi}), 
whose dispersion relations are given by 
$\hat \omega_s^2 \sim c_s^2 \hat k^2$ and $\hat \omega_f^2 \sim  \hat k^2+\hat{m}^2a^2 /\epsilon L^2$ in the small-scale limit, respectively.
Based on this observation, 
we separate the scalar perturbations into the ``slow" modes
 $\hat\phi_1^{(s)}$ with $\hat \omega_s^2 \sim c_s^2 \hat k^2$
and ``fast" modes $ \hat\phi_1^{(f)}$ with $\hat \omega_f^2 \sim  \hat k^2+\hat{m}^2a^2 /\epsilon L^2$, 
analogous to the splitting between background gravitational fields and gravitational waves.%
\footnote{The slow mode does not necessarily coincide with the adiabatic mode, as we will see in a few examples.}
Within this terminology, 
the adiabatic approximation (\ref{eq:adiabatic approx}) corresponds to a state with no excitation of the fast modes:
  \begin{align}
        \hat\phi_1^{(f)} \simeq 0 \,, \quad \hat\phi_1^{(s)} \simeq \hat\phi_1^{({\rm ad})} = \frac{1}{\hat{m}^2}\theta^l\frac{\partial \hat\zeta}{\partial \hat x} \,.
   \end{align}
Although we have defined the slow and fast modes here for the linear equations (\ref{app_zeta}) and (\ref{app_phi}), 
even at the non-linear level, 
the excitation of the fast modes will be a good indicator for the violation of the adiabatic approximation 
because of clear distinction of time scales between the slow and fast modes. 

\begin{figure}[t]
\centering
		\includegraphics[width=.8\linewidth]{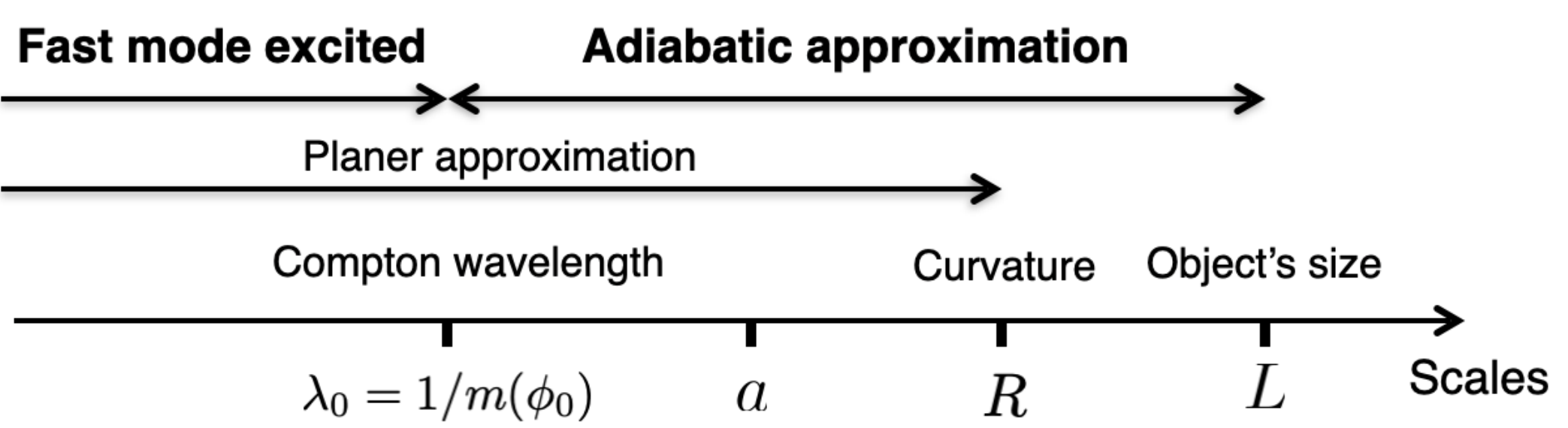}
\caption{
\baselineskip5mm
Hierarchy of the scales and the range where the adiabatic approximation is valid. 
Inhomogeneities with the length scales $k =\hat{k}/a > m(\phi_0)$ excite 
the scalar field around the potential minimum. 
The numerical analyses are performed within the region $R\leq r \leq R+a$, and $a$ gives the largest scale we 
consider in the analyses. 
}
\label{fig:scales}
\end{figure}

Before showing the results for the full non-linear analyses, 
in order to understand the fast mode excitation and the violation of the adiabatic approximation, 
let us analyze the linearized scalar equation in a dynamical background density. 
That is, we treat the density perturbation $\hat \rho_1$ as an external field ignoring the backreaction from the scalar field. 
With this simplification, we can analytically solve Eq.~(\ref{cham_phil})  by the Green's function method. 
We give a detailed analysis in Appendix \ref{massive_scalar}, and here we just refer to the results for the standing acoustic wave,
	\begin{align}\label{saw profile}
		\hat\rho_1(\hat{t},\hat{x}) = I \cos(\hat \omega_s \hat{t}) \cos(\hat k \hat{x}) 
	\end{align}
with $\hat \omega_s = c_s \hat k$, 
which describes the evolution of $\hat \rho_1$ in the linear regime. 
When we impose that the scalar field is initially given by the adiabatic profile
	\begin{align}\label{adiabatic ini main}
		\hat \phi_1(\hat t=0, \hat x) = -\frac{\hat \rho_1(\hat t=0,\hat x)}{\hat m^2} \,,\quad \pd_{\hat t} \hat \phi_1(\hat t=0, \hat x) = 0 \,,
	\end{align}
the solution is given by
	\begin{align}
		\hat \phi_1^{(s)}(\hat t, \hat x) &= -\left[ \frac{\hat k_{\rm ad}^2}{(1-c_s^2)\hat k^2+ \hat k_{\rm ad}^2} \right] \frac{\hat \rho_1}{\hat m^2} \,, \label{eq:ad slow} \\
		\hat \phi_1^{(f)}(\hat t, \hat x) &= \frac{I  \cos(\hat k \hat x)}{\hat m^2} \left[ \frac{ \hat k^2}{\hat k^2 + (1-c_s^2)^{-1}\hat k_{\rm ad}^2} \right] \cos(\hat \omega_f t) \label{eq:ad fast} \,,
	\end{align}
where $\hat \omega_f \equiv \sqrt{\hat k^2 + \hat k_{\rm ad}^2}$. 
Therefore, the adiabatic condition ($|\hat \phi_1^{(f)}| \ll |\hat \phi_1|$) is applicable if
    \begin{align}
        \hat k^2 \ll \hat{k}_{\rm ad}^2 = a^2 V''(\phi_0) \,,
    \end{align}
or equivalently
	\begin{align}
        		\frac{\hat k}{a} \ll m(\phi_0) \,,
         	\label{adiabatic_cond}
	\end{align}
taking into account $m^2(\phi_0) = V''_{\rm eff}(\phi_0) = V''(\phi_0)$ for a constant $\beta$. 
This result means that the adiabatic approximation can be violated if the scale of the perturbation $\hat \phi_1$ is smaller than the Compton wavelength $\lambda_0 = m^{-1}(\phi_0)$ (see Fig.~{\ref{fig:scales}}).

In the analysis above, 
we have assumed the adiabatic initial condition (\ref{adiabatic ini main}) to see how the adiabatic approximation is violated by the time evolution. 
However, 
when the non-linear term in the scalar equation is taken into account, 
the potential minimum for the potential $V(\phi)=\Lambda^4/\phi^{-1+1/\alpha}$ is determined as
    \begin{align}
         	 V'(\phi) = -\frac{\beta \rho}{M_{\rm pl}} \quad \Rightarrow \quad \hat{\phi} = \hat{\rho}^{-\alpha} \,,
        \label{eq:true minimum}
    \end{align}
and diverges in the low-density region $\hat{\rho} \simeq 0$. 
Meanwhile, as we have shown in Ref.~\cite{Nakamura2019} for a shell system, 
the adiabatic condition is not satisfied if the density profile is highly inhomogeneous even when the system is static.
Therefore, when the density profile has a large contrast, 
the scalar field does not trace the potential minimum (\ref{eq:true minimum}) because of the gradient term in Eq.~(\ref{cham_phi}).
From this consideration, in the non-linear analysis in the next subsection, we assume the following initial condition:
	\begin{align}\label{static ini linear}
		\hat \phi_1(\hat t=0, \hat x) = \hat \phi_s \,,\quad \pd_{\hat t} \hat \phi_1(\hat t=0, \hat x) = 0 \,,
	\end{align}
where $\hat \phi_s$ is a static solution for a given initial density profile. 
Under the same approximations to obtain the analytical results (\ref{eq:ad slow}), (\ref{eq:ad fast}), 
the static scalar profile is given by
	\begin{align}
	\label{static profile}
		\hat \phi_s = -\left( \frac{\hat k_{\rm ad}^2}{\hat k^2 + \hat k_{\rm ad}^2 } \right) \frac{\hat \rho_1}{\hat m^2} \,,
	\end{align}
and the solutions are given as
	\begin{align}
		\label{standing_wave_s}
		\hat\phi_1^{(s)}(\hat{t}, \hat{x}) &= -\left[ \frac{\hat k_{\rm ad}^2}{(1-c_s^2)\hat k^2 + \hat k_{\rm ad}^2 } \right] \frac{\hat \rho_1}{\hat m^2}, \\
		\label{standing_wave_f}
		\hat\phi_1^{(f)}(\hat{t}, \hat{x}) &=  -\frac{I \cos(\hat k \hat x)}{\hat{m}^2}  \left( \frac{c_s^2 \hat k^2}{k^2+k_{\rm ad}^2} \right) \left[ \frac{\hat k_{\rm ad}^2}{(1-c_s^2)\hat k^2 + \hat k_{\rm ad}^2 } \right]  \cos\left (\hat \omega_f \hat t \right). 
	\end{align}
It is noteworthy that the amplitude of Eq.~\eqref{standing_wave_f} takes the maximum value at the scale 
given by 
\b\label{eq:kres}
\hat k=\hat k_{\rm max} \equiv \left(1-c_s^2\right)^{-1/4}\hat k_{\rm ad}.
\e
Thus, 
taking into account the violation of the adiabatic approximation, 
we can  see the ratio of the fifth force to the gravitational force remains finite 
for large $\hat k/\hat k_{\rm ad}$ in the analysis with a linearized scalar field.
To examine how these results change due to the non-linear effect, we
will compare them with our non-linear numerical simulation in the next subsection.

\subsection{Numerical analysis of the non-linear system}
\label{numerical}
Here, we investigate how the fast mode $\phi^{(f)}$ is excited at the non-linear level and the resultant fifth force affects the fluid dynamics.  
In this subsection, we specify the potential as $V(\phi)=\Lambda^4/\phi^{2}$.
Thus, Eqs.~\eqref{nl_rho_1}--\eqref{nl_phi_1} can be written as follows: 
\begin{align}
\label{nl_rho_3}
\frac{\partial\hat\rho_1}{\partial \hat t}&=-\frac{\partial}{\partial \hat x}\left((\theta^l+\hat \rho_1)v\right), \\
\label{nl_euler_3}
\frac{\partial v}{\partial \hat t}&=-\frac{\partial}{\partial \hat x}\left[\frac{1}{2}v^2+\frac{c_s^2}{\gamma\theta}\frac{p_c}{\rho_c}\left\{(l+1)(\theta^l+\hat \rho_1)^{1/l}+\hat\Phi_{N1}+2\beta^2\epsilon\hat\phi_1\right\}\right], \\
\label{nl_poisson_3}
\frac{\partial^2\hat\Phi_{N1}}{\partial \hat x^2}&=\frac{a^2}{L^2}\hat\rho_1, \\
\label{nl_phi_3}
\frac{\partial^2\hat\phi_1}{\partial \hat t^2} &= 
\frac{\partial^2\hat\phi_1}{\partial \hat x^2} - 
	\frac{1}{\epsilon}\frac{a^2}{L^2}\left(\theta^{l}-\frac{1}{(\theta^{-l/3}+\hat\phi_1)^3}+\hat\rho_1\right),
\end{align}
where $v=\partial\hat\zeta/\partial \hat t$. 
We numerically solve the above equations in the region $0\leq\hat x\leq1$ 
imposing the periodic boundary condition. 
The integrability condition of Eq.~\eqref{nl_poisson_3} gives the following condition for the density profile:
\b
\int^1_0\hat\rho_1~\dd \hat x=0.
\label{period_N}
\e
Here we consider the following initial conditions for the numerical simulation:
\b
\label{nonlinear_ini}
\hat\rho_1|_{t=0}=-I\cos(2\pi \hat x), ~~\hat\phi_1=\hat\phi_{\rm s}(\hat x), ~~v|_{ t=0}=\partial_{\hat t} \phi|_{t=0}=0, 
\e
where $\hat \phi_{\rm s}(\hat x)$ is the static profile obtained by numerically solving the scalar field equation \eqref{nl_phi_3} with 
the density profile fixed to the initial profile. 
Note that 
the density profile satisfies Eq.~\eqref{period_N} and 
the amplitude $I$ should be smaller than $\rho_0/\rho_c=\theta^l$ 
so that the total density is positive.
We use a central scheme with MUSCL method (Mono Upstream-centered Scheme for Conservation Laws) for the fluid part.
The scalar field equation
and the Poisson equation are solved by 
the 4th order Runge-Kutta method.

We set $\beta^2=1/2$ and the polytropic index as $\gamma=2$ in the following calculations.
In the local region, the parameter $\theta$ is approximately constant and we can obtain the same equations with redefining the quantities as $\hat{\rho}_1 \to \theta^l\hat{\rho}_1$, $\hat{\Phi}_{N1} \to \theta\hat{\Phi}_{N1}$, $\hat{\phi}_1 \to \theta^{-l/3}\hat{\phi}_1$, 
$\epsilon \to \theta^{1+3/l}\epsilon$, and $(a^2/L^2)\to \theta^{1-l} a^2/L^2$. Therefore, we set $\theta =1$ in the following analysis.
Then there are four parameters, $I$, $\epsilon$, $a/L$, and $c_{s}^{2}$.

\subsubsection{Evolution of the scalar field profile}
\label{sec:general behavior}

\begin{figure}[t]
\centering
\includegraphics[width=16cm]{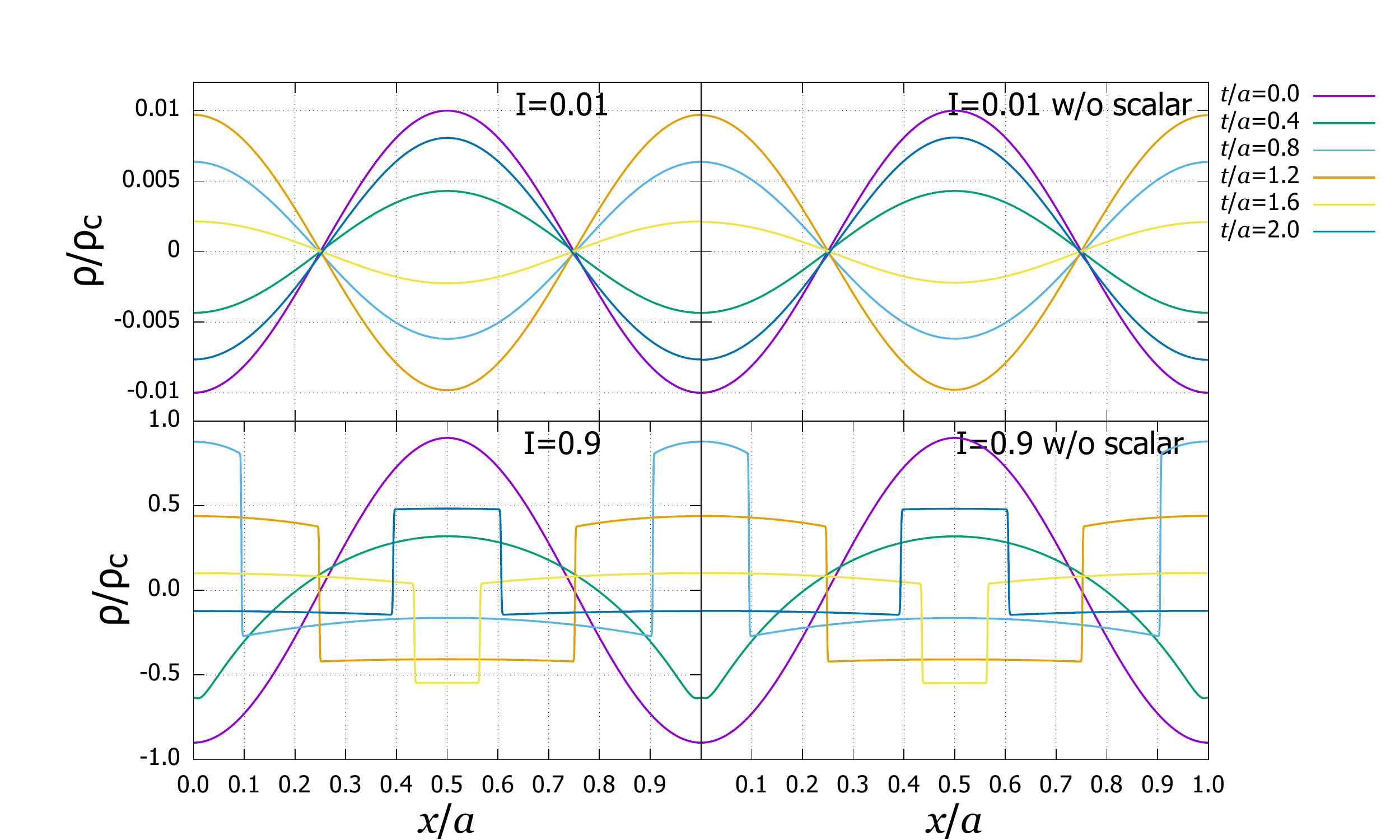}
\centering
\caption{
\baselineskip5mm
Evolution of the density profile with the scalar field (left) and without the scalar field (right). }
\label{density}
\end{figure}

\begin{figure}[t]
\centering
\includegraphics[width=17cm]{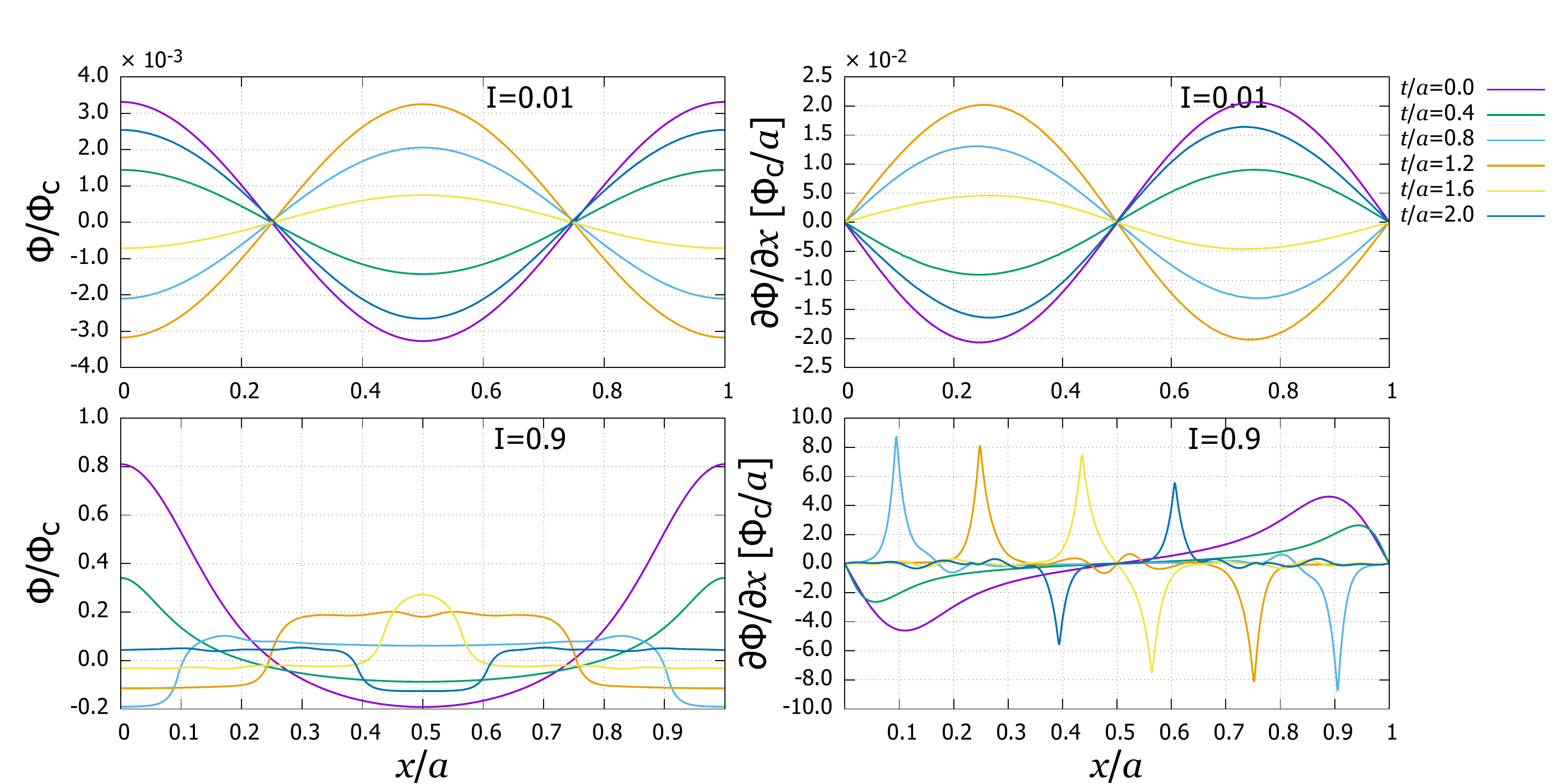}
\centering
\caption{
\baselineskip5mm
Evolution of the scalar field profile (left) and its spatial derivative (right). }
\label{scalar}
\end{figure}

First, we show the behavior of the density (Fig.~\ref{density}) and the scalar field (Fig.~\ref{scalar}) 
with $\epsilon=10^{-4}, a^2/L^2=0.1, c_s^2/\gamma=0.1$ 
for each value of the initial amplitude $I$. 
In the left panels of Fig.~\ref{density}, we show the evolution 
of the density perturbation for $I=0.01$ (top) and $I=0.9$ (bottom). 
To examine the effect of the scalar field on the evolution of the density, 
in the right panels we show the results for the same parameters with the scalar field turned off, i.e., removing the $\hat \phi_1$ term from Eq.~(\ref{nl_euler_3}).
For $I=0.01$, the density perturbation can be well approximated by the linear 
standing-wave solution given in Eq.~\eqref{saw profile}. 
For $I=0.9$, 
we can find a shock-wave structure 
due to the non-linearity of the fluid equation as shown in Fig.~\ref{density}. 
Even in such a case, the scalar field does not significantly affect the evolution of the density perturbation. 

\begin{figure}[t]
\centering
		\includegraphics[width=17cm]{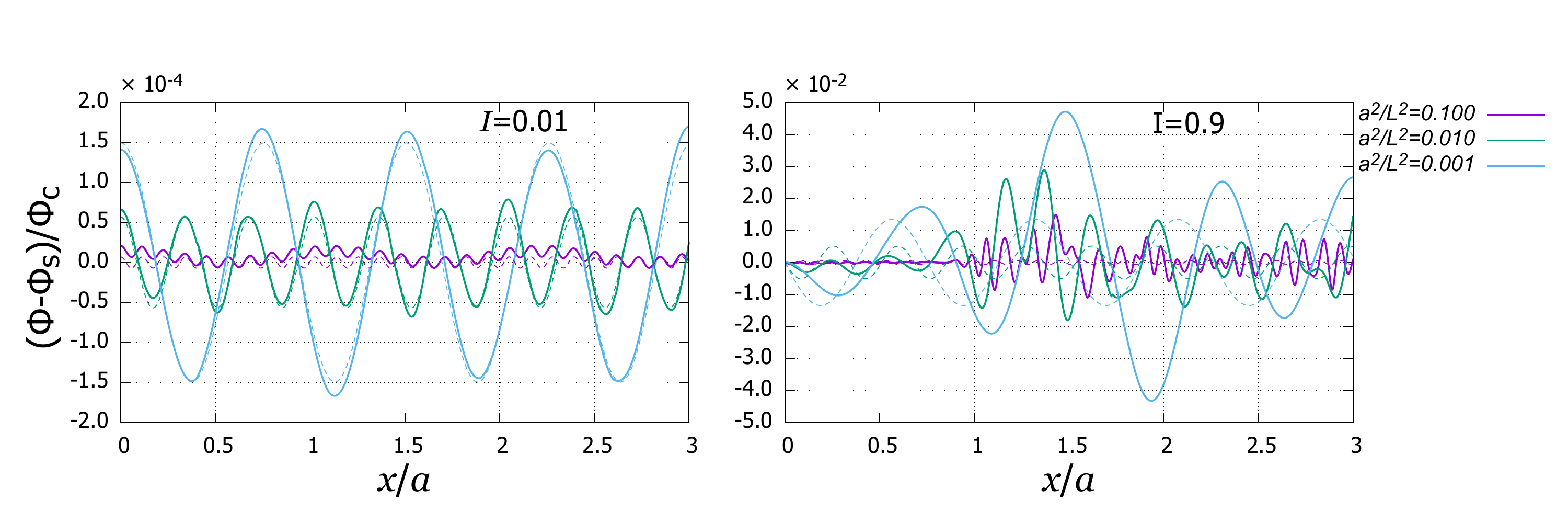}
\centering
\caption{
\baselineskip5mm
The fast mode amplitude for $I=0.01, 0.9$ at $\hat x=0.4$. 
The dashed lines represent the analytical estimation given by Eq.~\eqref{standing_wave_f}.
}
\label{phi_comp}
\end{figure}

Let us check the evolution of 
the scalar field.  
In Fig.~\ref{scalar}, we show the evolution of the profile of the scalar field (left) 
and its spatial derivative (right). 
For $I=0.01$, we find a standing wave behavior as expected from the linear analysis, while 
the profile is significantly different from the cosine function for $I=0.9$.  
In Fig.~\ref{phi_comp}, 
we also show how the amplitude evolves at the position $\hat x = 0.4$ for the small ($I=0.01$) and large ($I=0.9$) density perturbations,
subtracting the mode slowly varying with the density perturbations. 
When the amplitude $I$ is small, we can identify the subtracted mode as the slow mode (\ref{standing_wave_s}) in the linear analysis.  
Therefore, we call the subtracted and remaining modes as the slow mode $\phi_1^{(s)}$ and the fast mode $\hat{\phi}_1^{(f)}$, respectively. 
In the non-linear regime, the linear solution (\ref{standing_wave_s}) is no longer valid. 
Hence, as a proxy for the slow mode $\phi_1^{(s)}$, we subtract the static solution $\hat{\phi}_s$ although they are slightly different as we can see it from the difference between Eqs.~\eqref{static profile} and \eqref{standing_wave_s}.
We also show the linear solution of the fast mode \eqref{standing_wave_f} in Fig.~\ref{phi_comp}.
We find that the analytical estimation \eqref{standing_wave_f} gives a good approximation for the small amplitude case, i.e.\ $I=0.01$ irrespective 
of the value of $a^2/L^2$. 
On the other hand,  
the fast-mode amplitude is larger than the analytical estimation for $I=0.9$, 
where the non-linear effect is significant. 
We also find similar behaviors for different values of $a^2/L^2$ as shown in Fig.~\ref{phi_comp}.

\begin{figure}[t]
\centering
		\includegraphics[width=8cm]{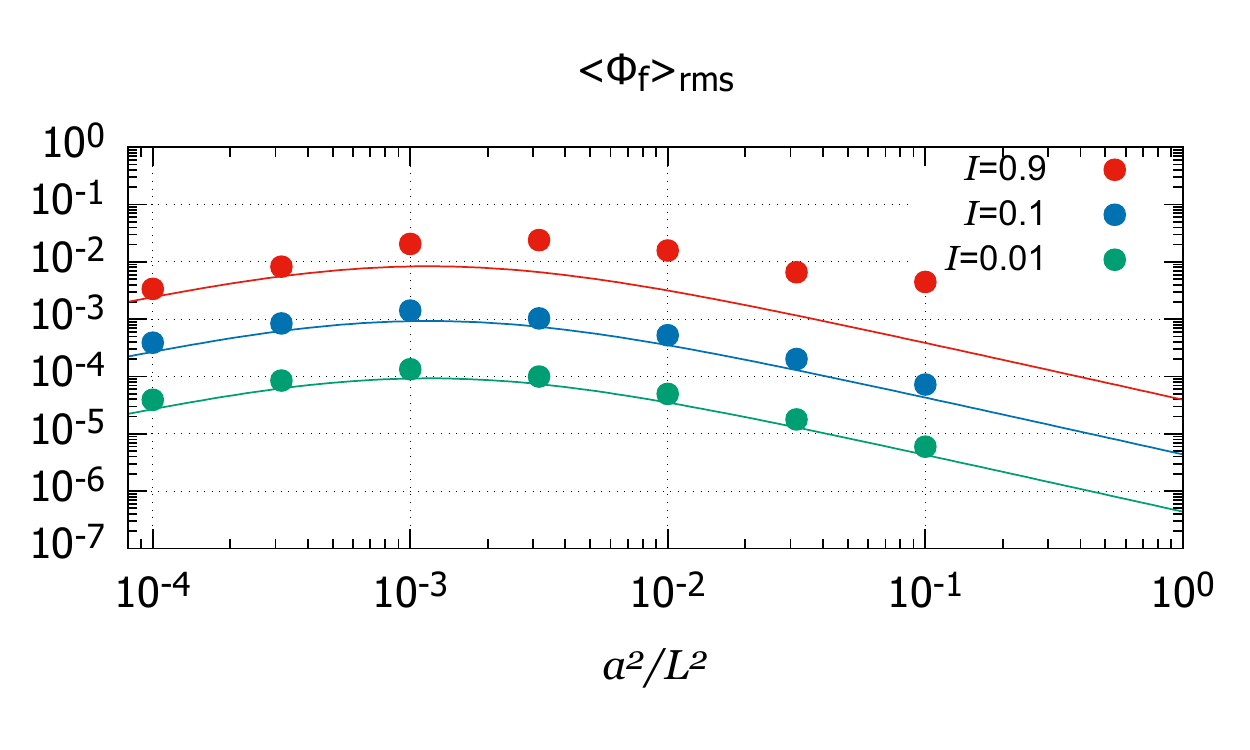}~
		\includegraphics[width=8cm]{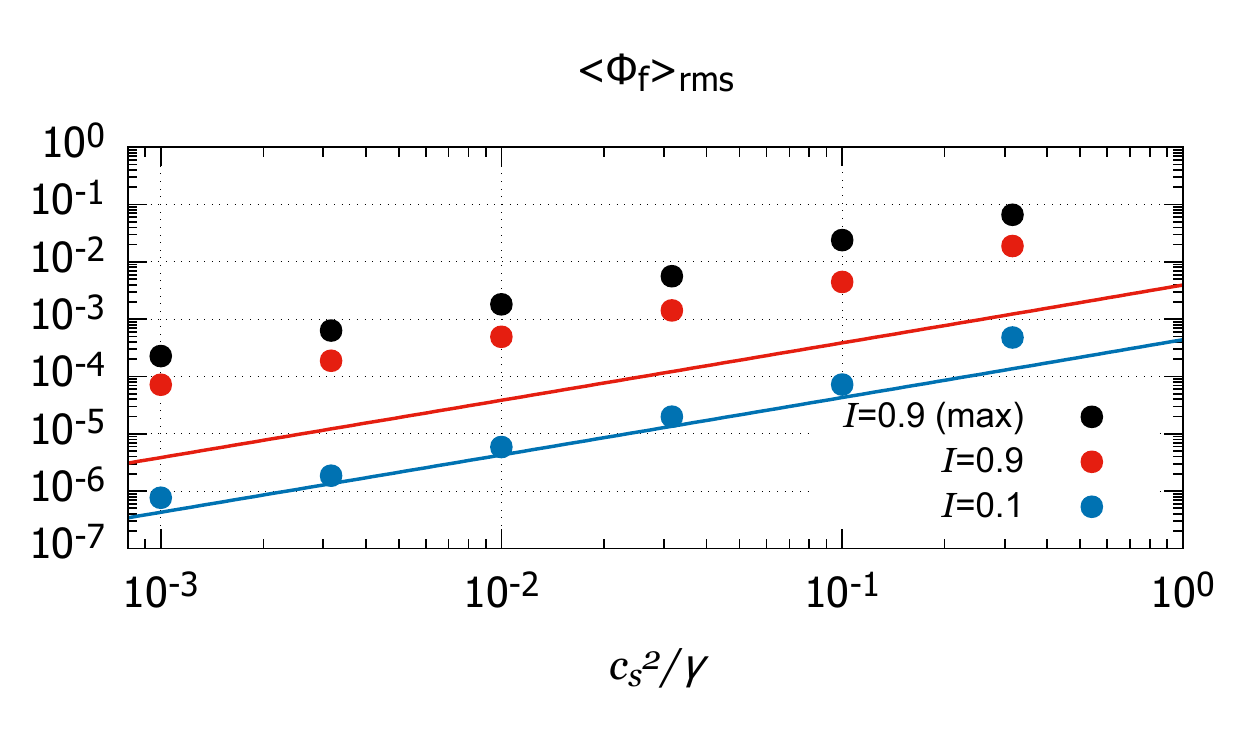}
\\
		\includegraphics[width=8cm]{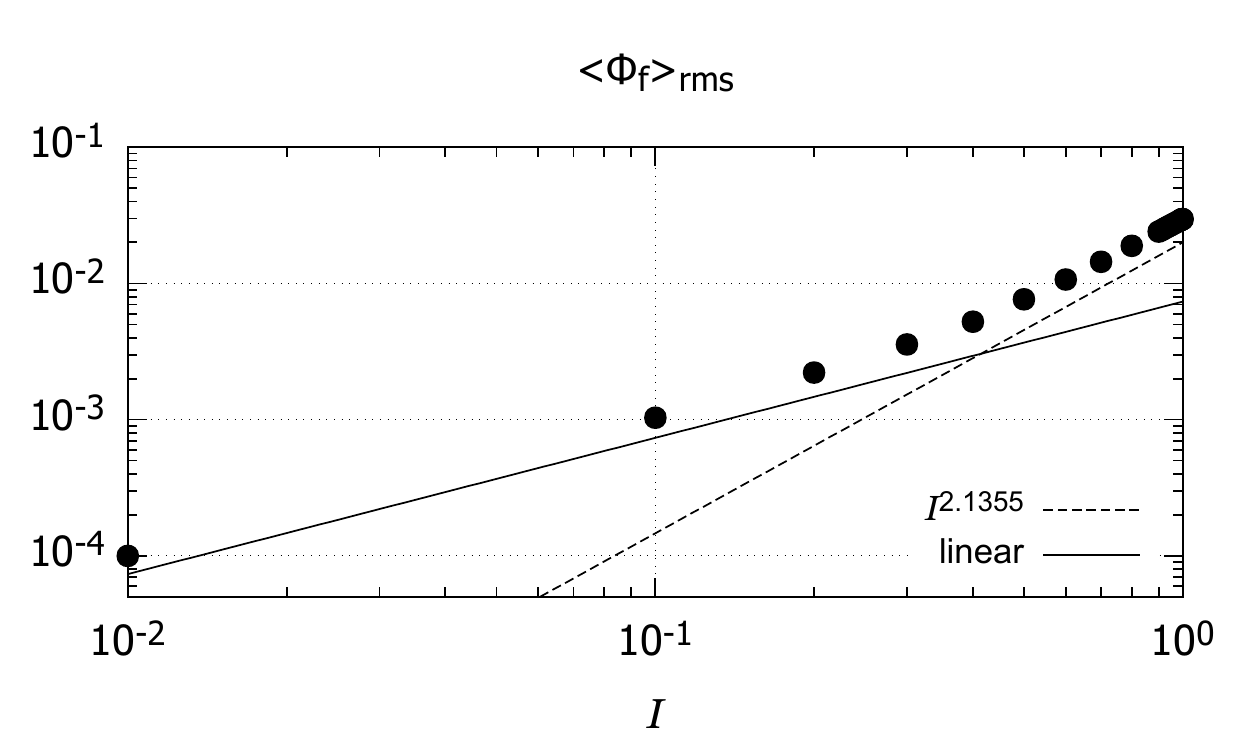}
\centering
\caption{
\baselineskip5mm
Parameter dependence of the root-mean-square amplitude of the fast mode $\av$. 
In the top panels,
the dots show the numerical results and
the solid lines show the estimation from the linear solution \eqref{standing_wave_f}.
The top-left panel shows $\av$ for $c_s^2/\gamma=0.1$ with $I$ and $a^2/L^2$ varied, 
while the top-right panel for $a^2/L^2=0.1$ with $I$ and $c_s^2/\gamma$ varied.
In the top-right panel, the dots for ``$I=0.9\,(\rm max)$'' represent $\av$ for $a^2/L^2=10^{-5/2}$, for which $\langle \hat\phi_f \rangle_{\rm rms}$ approaches the maximum value as is shown in the top-left panel. 
The bottom panel shows the dependence on $I$ for $a^2/L^2=10^{-5/2}$. 
The linear analysis predicts that $\av$ linearly depends on $I$, and it matches the numerical result well. However, it exhibits $I^{2.1355}$ dependence for higher $I$. 
This shows the breakdown of the linear dependence due to the non-linearity, while the value of exponent may be dependent on the choice of $\hat \phi_1^{(s)}$ used in the numerical analysis.}
\label{para_dep}
\end{figure}

To investigate further when the deviation from the linear solution (\ref{standing_wave_f}) becomes important, 
let us estimate the root-mean-square of the fast mode amplitude\footnote{\baselineskip5.5mm Note that we subtract the static profile $\hat\phi_s$ from the scalar field amplitude because $\hat\phi^{(s)}$ cannot be estimated exactly for large $I$ cases as we mentioned in the previous paragraph.
Consequently, we named the averaged value as $\av$ instead of using $\hat\phi^{(f)}$. }
$\av:=\langle(\hat \phi_1-\hat \phi_s)^2\rangle^{1/2}$
 for various values of the parameters $I$, $a/L$, and $c_{s}^{2}$. 
Here, the averaging is performed over the whole spatial domain ($0\leq\hat x\leq 1$) and one unit of acoustic time scale ($0\leq \hat t\leq 1/\sqrt{c_s}$).
The top panels of Fig.~\ref{para_dep} show dependence of $\av$ on $a^2/L^2$ and $c_s^2/\gamma$ 
for each value of the initial amplitude $I$,
while the bottom panel shows dependence on the initial amplitude $I$.

We can see that the analytical estimation \eqref{standing_wave_f} reproduces the numerical results of $\langle \hat\phi_f \rangle_{\rm rms}$ accurately 
when the initial amplitude is small, while the numerical results tend to be slightly larger than the analytical results for any parameter. 
The appearance of a peak of $\langle \hat\phi_f \rangle_{\rm rms}$ around $a^2/L^2 \simeq 10^{-5/2}$ is expected from the linear analysis \eqref{standing_wave_f}: 
it corresponds to the maximum condition~\eqref{eq:kres} 
with $\hat{k}=2\pi$, which is specified by the initial configuration~\eqref{nonlinear_ini}. 
The discrepancy of $\langle \hat\phi_f \rangle_{\rm rms}$ from the analytical estimate arises probably because the numerically obtained $\hat{\phi}_f$ is slightly contaminated by the slow mode~(\ref{standing_wave_s}). 
As we commented above, 
we use the static solution as $\hat \phi_1^{(s)}$ at each moment in the numerical calculation for $\av$. 
This choice is useful but does not completely remove the slow mode component from $\hat\phi_1$, 
and the remaining slow mode gives an additional contribution to $\av$.

For larger initial amplitude $I>0.1$, 
$\av$ tends to be larger than the analytical estimation as shown in Fig.~\ref{para_dep}. 
This enhancement may be qualitatively understood by considering the 
spatial dependence of the effective scalar field mass.%
\footnote{
\baselineskip5mm
One might expect the shock wave structure can be the origin of the enhancement of the scalar field amplitude. 
In order to check this possibility, 
we derive the linearized scalar field behavior 
by considering the general density profile (see Appendix \ref{massive_scalar}).
Then we find that the fast mode amplitude for general matter profile Eq.~\eqref{general_f} 
cannot explain the parameter dependence in Fig.~\ref{para_dep}, e.g.\ $a^2/L^2$ dependence. 
Therefore we discard this possibility. 
}
In the analysis with the linearized scalar field equation, 
we fixed the effective mass as that in the background density: $m^2 = m^2(\phi_0)$. 
However, 
when the density perturbation has a large initial amplitude $I$, 
the scalar field mass can largely deviate from that in the background density depending on the local value of the density perturbation. 
From Eq.~\eqref{mass}, we obtain $m^2\propto\rho^{4/3}$, and 
the $\rho$ dependence of the fast mode can be roughly estimated as $\hat\phi_1^{(f)}\propto(1+\hat\rho_1)^{-4/3}$ 
through Eq.~\eqref{standing_wave_f}. 
Because of the inverse-power dependence on the density perturbation satisfying Eq.~\eqref{period_N},
the enhancement effect in the lower density region due to smaller mass is more effective 
than the reduction effect in the higher density region. 
As a consequence, the enhancement of the amplitude would be observed on average. 
For a sufficiently small value of $a^2/L^2$, 
the typical wavelength of the fast mode becomes much longer than 
that of the density perturbation so that the scalar field is not affected by 
the inhomogeneity. 
Then we expect that the effective mass of the scalar field can be approximated by the background value for 
a sufficiently small value of $a^2/L^2$ even for the large value of the initial amplitude $I$. 
Actually, 
the fast mode amplitude shows good agreement with the analytic result with the fixed scalar field mass 
when $a^2/L^2$ is smaller than the value corresponding to $\hat k=\hat k_{\rm max}$.

Let us consider an extreme case where the initial amplitude is so large that 
an underdense region becomes nearly vacuum. 
Although the scalar-field potential does not have a stable point in the vacuum region, 
as is shown in Ref.~\cite{Nakamura2019} for static configurations, 
there always exists an upper bound on the scalar field value in a finite spatial region. 
In our dynamical calculation, 
the fast mode defined as the deviation from the static configuration 
is always finite as shown in Fig.~\ref{para_dep}~(see Appendix~\ref{st_profile} for the difference between adiabatic profile and the static profile). 
This result is consistent with the results in Ref.~\cite{Nakamura2019}. 

\subsubsection{Fifth force}
\label{sec:comparison}

\begin{figure}[t]
\centering
		\includegraphics[width=16cm]{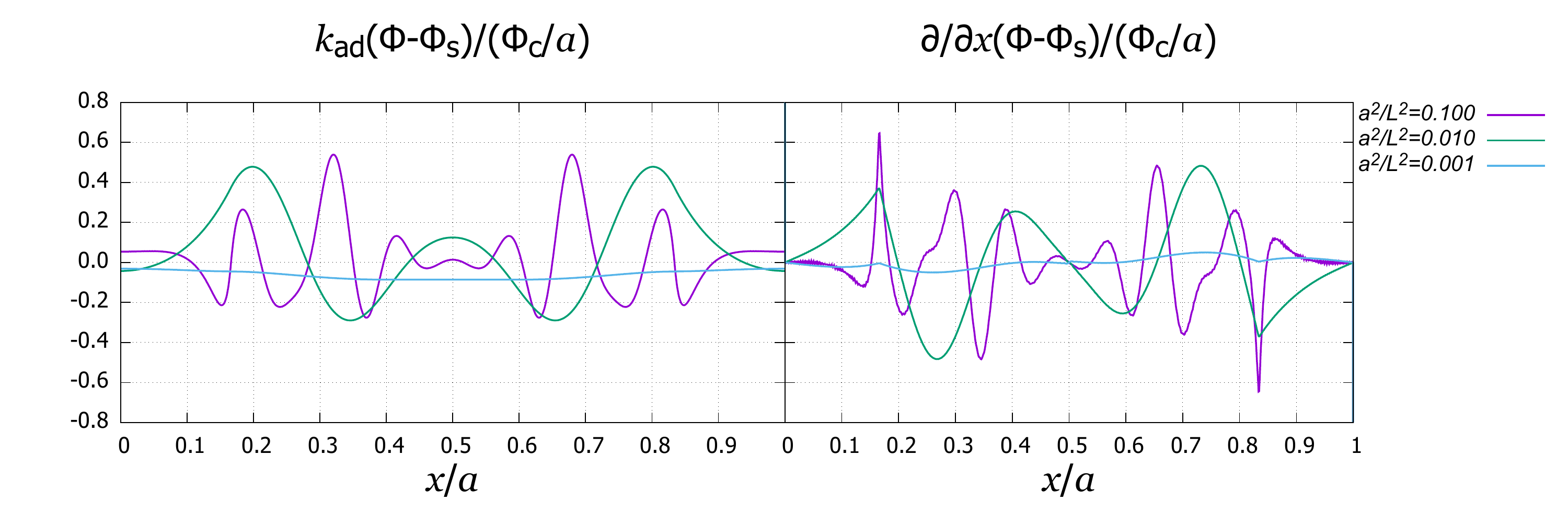}
\centering
\caption{
\baselineskip5mm
The fast mode profile (left) and its spatial derivative (right) for $I=0.9, c_s^2/\gamma=0.1$ at $\hat{t}=1.0$. }
\label{fifth_comp}
\end{figure}

\begin{figure}[t]
\centering
		\includegraphics[width=8cm]{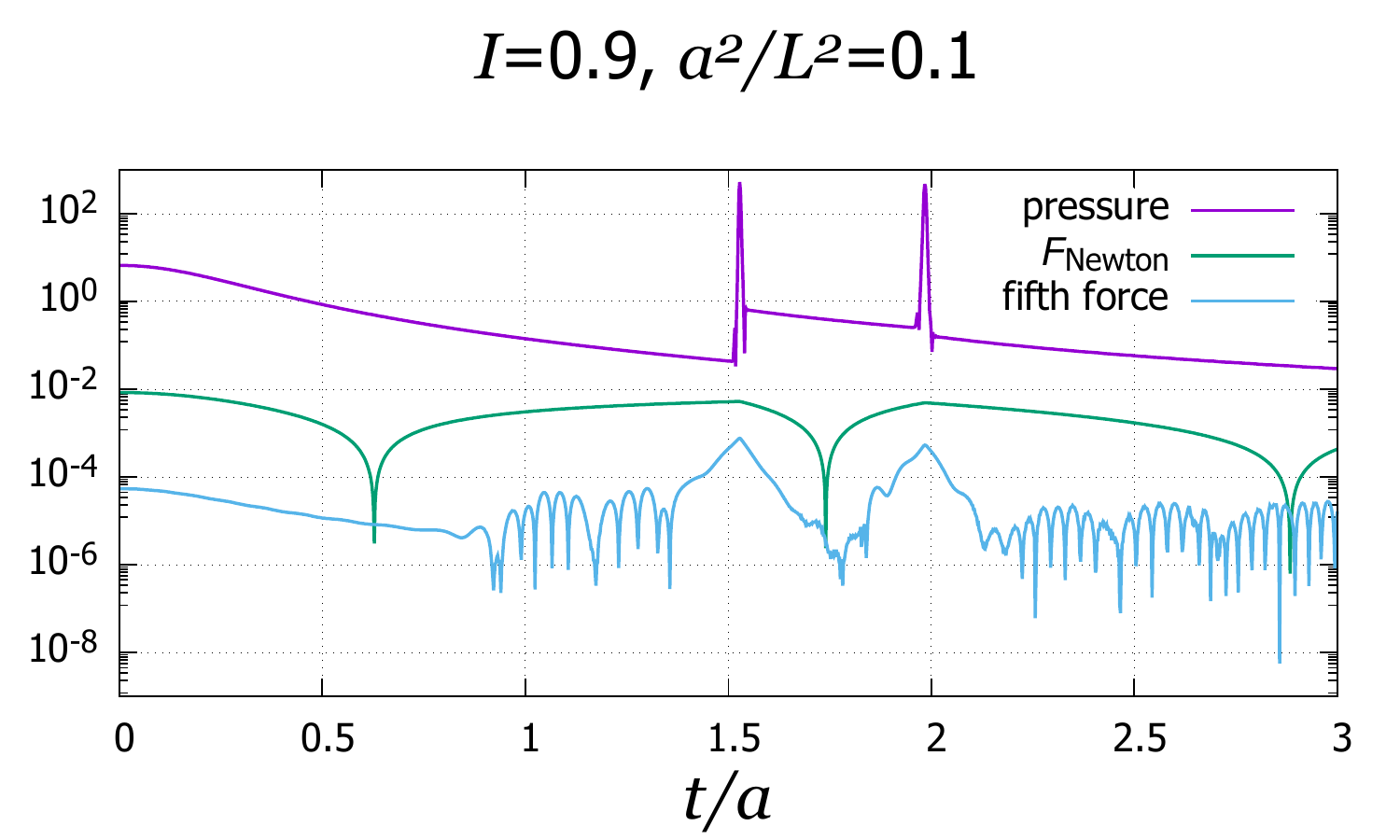}~
		\includegraphics[width=8cm]{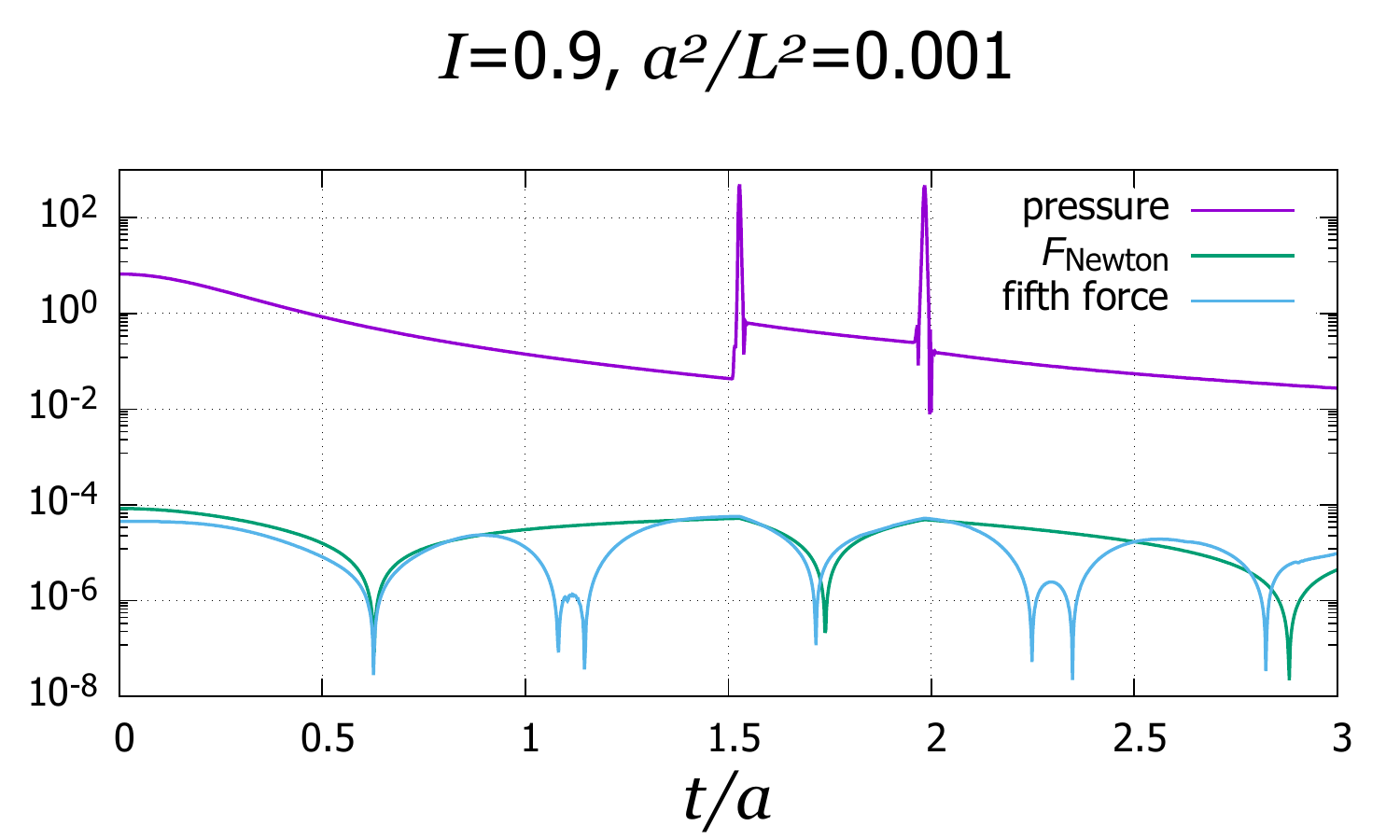}
\centering
\caption{
\baselineskip5mm
Time evolution of each term in the right-hand side of Euler equation~(\ref{nl_euler_3}) at $\hat x=0.4$,
where we plot
$(l+1)\partial_{\hat x}(\theta^l+\hat\rho_1)^{1/l}$ as the pressure, $\partial_{\hat x}\hat\Phi_{N1}$ as $F_{\rm Newton}$, and $\epsilon\partial\hat\phi_1/\partial\hat x$ as the fifth force.
The left figure shows the case of $I=0.9$, $a^2/L^2=0.1$ and the right figure shows $I=0.9$, $a^2/L^2=0.001$.}
\label{euler}
\end{figure}

Next, we see how large the fifth force is enhanced compared to the Newtonian force.
We consider only the case where $\hat k<\hat k_{\rm ad}$ since we are interested in the non-linear effect.
In order to give an estimate for the fifth force, we need to know the 
typical spatial scale of the fast mode configuration. 
We plot the spatial configuration of the fast mode and the fifth force at $\hat t=1$ 
in Fig.~\ref{fifth_comp}.
Comparing the left and right panels, 
we find that the spatial derivative of the fast mode is roughly given by multiplying 
$\hat k_{\rm ad}\sim 1/\sqrt{\epsilon}\cdot a/L$ to the amplitude.
This shows that the typical spatial scale of the fast mode is given by $1/\hat k_{\rm ad}$. 
Since the wavenumber of the initial density perturbation is $\hat k=2\pi$, 
modes with $\hat k\sim \hat k_{\rm ad}\gg2\pi$ are generated through the non-linear terms. 
We note that, although modes with $\hat k<\hat k_{\rm ad}$ may be 
also generated by the non-linear effect, their contribution to the fifth force is smaller than that of $\hat k\sim \hat k_{\rm ad}$. 
In order to see the effect of the fifth force on the dynamics, 
in Fig.~\ref{euler}, we show the time evolution of the Newtonian force and the fifth force as well as the pressure gradient force in the Euler equation \eqref{nl_euler_3}. 
We find that the ratio between the Newtonian and the fifth forces becomes order of unity when $a^2/L^2 = 0.001$.  
Nevertheless, 
the pressure gradient term is always dominant in the time evolution,
which explains why the scalar does not affect the fluid dynamics as shown in Fig.~\ref{density}.

Finally, let us try to understand the above result more quantitatively.
As we saw in Fig.~\ref{fifth_comp}, spatial derivative of the scalar field can be estimated by assuming that 
$\hat k_{\rm ad}$ mode is dominant, while we can use $\hat k$ to estimate the pressure gradient and the Newtonian force.  
Therefore, each ratio of the forces can be 
roughly estimated as
\begin{align}
						\frac{F_{\phi}}{F_{\rm Newton}}
	&\sim\frac{\epsilon \hat{k}_{\rm ad} |\hat\phi_1^{(f)}|}{1/\hat k \cdot (a^2/L^2) \hat\rho_1}
	\sim \sqrt{\epsilon}\left(\frac{L}{a}\right) \cdot \hat{k}\frac{|\hat\phi_1^{(f)}|}{\hat{\rho}_1}  \sim  \frac{\hat{k}}{\hat{k}_{\rm ad}} \frac{|\hat\phi_1^{(f)}|}{\hat\rho_1} \,, \\
\frac{F_{\phi}}{F_{\rm pressure}}
\label{eq:ratioNF}
&\sim
~
\left(\frac{a}{L}\right)^2\frac{F_{\phi}}{F_{\rm Newton}}
~
\sim\sqrt{\epsilon}\left(\frac{a}{L}\right)\cdot \hat{k}\frac{|\hat\phi_1^{(f)}|}{\hat{\rho}_1}  
\,
\sim  \left(\frac{a}{L}\right)^2\frac{\hat{k}}{\hat{k}_{\rm ad}} \frac{|\hat\phi_1^{(f)}|}{\hat\rho_1} \,. 
\end{align}
Focusing on the $a/L$ dependence, we expect the ratio between the Newtonian force and fifth force 
can be amplified 
for small scale perturbations in addition to the enhancement of $|\hat\phi_1^{(f)}|$ due to the non-linearity. 
In fact, the ratio becomes order of unity for $a^2/L^2=0.001$, where $\hat k\sim \hat k_{\rm ad}$, 
although the ratio to the pressure gradient is tiny in any case. 
In order to see the significance of the enhancement of the scalar field amplitude $|\hat\phi_1^{(f)}|$
due to the non-linear effect, let us estimate Eq.~\eqref{eq:ratioNF} using the fast mode amplitude \eqref{standing_wave_f} derived from the linearized scalar field equation.  
For $\hat k\lesssim\hat k_{\rm ad}$, we obtain 
$F_{\phi}/F_{\rm Newton} \sim
	c_s^2\left(\hat{k}/\hat{k}_{\rm ad}\right)^3\lesssim c_s^2$. 
Since $c_s^2$ is always smaller than $1$, 
the further amplification to order of unity would be due to
the non-linear enhancement of $|\hat\phi_1^{(f)}|$.

It is also noteworthy that the relative amplitude of the pressure gradient force to the other forces 
is not reduced even for $c_s^2 \to 0$: 
the parameter dependence of the relative amplitude is determined by $b^{-1}$ (see Eq.~\eqref{nl_euler_1}), but 
it is given by $b^{-1} = \xi_s^{-2}/(l+1)$ and cannot be much smaller than unity. 
Since $\xi_s$ is obtained by solving the background equilibrium condition \eqref{m_lane_emden}, 
this fact can be interpreted as a result of the balance between the pressure gradient force and the others at the background level.


\section{Conclusion}
\label{conclusion}
We analyzed the dynamics of the chameleon field in a simple stellar model 
with the polytropic equation of state and investigated the significance of the fifth force. 
First, we derived the exact dynamical equations both for the chameleon field and matter in the Newtonian limit. 
With these equations, we studied two aspects of the system: stability and dynamical excitation of the fifth force. 
As for the first issue, 
we performed the linear analysis of spherical modes with  the adiabatic approximation, and 
confirmed that the chameleon field does not significantly affect the stability of the stellar object
for a small value of a screening parameter. 
We addressed the second issue particularly
focusing on the importance of the chameleon-field dynamics and its backreaction to the matter. 
Analytically solving the linear equations with the planer approximation, 
we  showed that the adiabatic approximation is violated on scales smaller than the chameleon field's Compton wavelength, and 
then the dynamics of the chameleon field can be important. 
Next, we numerically analyzed the dynamical equations 
taking all non-linear terms into account. 
We found that 
the qualitative behavior of the chameleon field can be understood from the solutions of the linearized scalar field equations,  
but the amplitude is more enhanced in the non-linear regime. 

It should be emphasized that, according to our analysis, 
the chameleon field can be excited and the fifth force can be comparable to the Newtonian force
even when the object is well screened for the averaged density configurations. 
It implies that the fifth force can be comparable to the Newtonian force inside the object
even when the fifth force is not detected outside the object.
We also studied whether the enhanced fifth force can affect the dynamics of the total system. 
In the stellar model, we found that the pressure gradient force is always dominant and the fifth force is irrelevant. 
One reason for this suppression of the fifth force is 
the static equilibrium condition imposed on parameters for the background stellar system. 
If we consider a non-equilibrium system, 
the fifth force might significantly affect the dynamics of the total system. 
It is also noteworthy that the excited chameleon field is a 
dynamical mode, which can propagate outwards and may leak from the surface of the star \cite{Silvestri2011}. 
If the amplitude of the chameleon field is sufficiently large and the leakage is significant,  
 it would be detectable as a scalar mode of gravitational waves. 
Our result showed that  
the flux of the fast mode is estimated as $\sim\rho_cc_s(\hat\phi_1^{(f)})^2$, and is not necessarily highly suppressed even 
when the star is well screened on average. 
It would be interesting to investigate how much the excited mode can leak 
from the surface of the star and its detectability. 

In Ref.~\cite{Haar2020}, a similar result was reported for the kinetic screening mechanism (the K-Mouflage model): 
the fifth force does not affect the dynamics of the matter unless the gravitational collapse happens. 
In our case, we chose the polytropic index so that the background solution can be stable and 
focused on the situation where the gravity term does not become dominant,  that is, no gravitational collapse happens.  
It would be interesting to investigate a similar system for other screening mechanisms. 
In particular, our scheme can be straightforwardly extended to  symmetron and environmental dilaton. 
We leave them for future works. 


\acknowledgements
This work was
supported by 
JSPS KAKENHI Grants No. JP19H01895 (C.Y.), No. 18K03623 (N.T.), No. 17K14286, No. 19H01891(R.S.)
the European Union's H2020 ERC Consolidator Grant ``Matter and strong-field gravity: New frontiers in Einstein's theory'' grant agreement no.\ MaGRaTh- 646597, the H2020-MSCA-RISE-2015 Grant No.\ StronGrHEP-690904 and the European Union's H2020 ERC, Starting
Grant Agreement No.~DarkGRA--757480. (T.I.).

\begin{appendix}
\section{Linear analysis for the fast mode excitation}
\label{massive_scalar}

In this appendix, we analyze the linear scalar equation (\ref{cham_phil}) 
    \begin{align}
        \left(-\frac{\partial ^2}{\partial \hat t^2}+\frac{\partial ^2}{\partial \hat x^2}\right)\hat\phi_1 = \frac{1}{\epsilon}\frac{a^2}{L^2}\left(\hat m^2\hat\phi_1+\hat\rho_1\right),
    \end{align}
with an emphasis on a condition that the fast mode is excited, i.e.\ the adiabatic approximation becomes invalid.
Here, the mass $\hat m$ is evaluated for the background scalar $\phi_0$ and constant. 
Moreover, throughout this section, we treat the density perturbation $\hat{\rho}_1$ as an external field by neglecting the backreaction from the scalar field $\hat{\phi}_1$ to its dynamics.

\subsection{General solution}
The Green's function for the linear scalar equation (\ref{cham_phil}) is given by
	\begin{align}
		G(\hat{t},\hat{x}) = \sum_{n \in \iz} \frac{\sin(\omega_n \hat{t})\theta(\hat{t})e^{ik_n \hat{x}}}{\omega_n} \,,
	\end{align}
where the wavenumber and frequency are restricted to
    \begin{align}\label{eq:periodic cond}
		 \hat k_n \equiv 2\pi n ~ (n \in \iz) \,, \quad \hat\omega_n \equiv \sqrt{\hat k_n^2 + \hat k_{\rm ad}^2} 
	\end{align}
with $k_{\rm ad} \equiv a\hat m/(\sqrt{\epsilon}L)$, 
because we have imposed the periodic boundary condition $G(\hat{t},\hat{x}+1)=G(\hat{t},\hat{x}$). 
Therefore, an analytical solution for a general density perturbation $\hat{\rho}_1$ can be obtained as
    \begin{align}\label{eq:green sol}
		\hat{\phi}_1(\hat{t}, \hat{x}) - \hat{\phi}_{\rm hom}(\hat{t},\hat{x}) &= 
		- \frac{1}{\epsilon} \frac{a^2}{L^2} \int {\rm d}\hat{t}' \int {\rm d}\hat{x}' ~ G(\hat{t}-\hat{t}',\hat{x}-\hat{x}') \hat{\rho}_1(\hat{t}',\hat{x}') \\
		&=- \frac{1}{\epsilon} \frac{a^2}{L^2} \sum_{n \in \iz} \left\{ \frac{e^{i\hat k_n \hat{x}}}{\hat\omega_n} \int_{0}^{\hat{t}} {\rm d}\hat{t}' \int_{-1/2}^{1/2} {\rm d}\hat{x}' \sin[\hat\omega_n(\hat{t}-\hat{t}')]e^{-i\hat k_n \hat{x}'}\hat{\rho}_1(\hat{t}',\hat{x}') \right\} \,,
	\end{align}
where $\hat{\phi}_{\rm hom}$ is a homogeneous solution and to be determined so that $\hat{\phi}_1$ will satisfy an appropriate initial condition. 

\subsection{Standing acoustic wave}

In this section,
we give results for a standing acoustic wave (\ref{saw profile}),
	\begin{align}
		\hat\rho_1(\hat{t},\hat{x}) = I \cos(\hat \omega_s \hat{t}) \cos(\hat k_{n_\ast} \hat{x}) \quad (\hat \omega_s = c_s \hat k_{n_\ast}) \,.
	\end{align}
Substituting this density profile to Eq.~(\ref{eq:green sol}), 
we find
	\begin{align}
		\hat \phi_1(\hat t, \hat x) - \hat \phi_{\rm hom}(\hat t, \hat x) 
		= \frac{I}{m^2}  \frac{\hat \omega_0^2}{\hat \omega_s^2 - \hat \omega_{n_\ast}^2} \cos(\hat k_{n_\ast} \hat x) \left[ \cos(\hat \omega_s \hat t) - \cos(\hat \omega_{n_\ast} \hat t) \right] \,.
	\end{align}
To determine the homogeneous solution $\hat \phi_{\rm hom}(\hat t, \hat x)$, 
we consider two types of initial conditions.

\subsubsection{Adiabatic initial condition}

To see how the adiabatic approximation is violated as time evolves, 
we impose that the scalar field is initially given by the adiabatic profile, 
	\begin{align}\label{adiabatic ini}
		\hat \phi_1(\hat t=0, \hat x) = -\frac{\hat \rho_1(\hat t=0,\hat x)}{\hat m^2} \,,\quad \pd_{\hat t} \hat \phi_1(\hat t=0, \hat x) = 0 \,.
	\end{align}
Then, the homogeneous solution is determined as
	\begin{align}
		\hat \phi_{\rm hom}(\hat t, \hat x) = -\frac{I\cos(\hat k_{n_\ast} \hat x)}{\hat m^2}\cos(\hat \omega_{n_\ast} \hat t) \,,
	\end{align}
and the slow and fast modes are given by
	\begin{align}
		\hat \phi_1^{(s)}(\hat t, \hat x) &= \frac{I}{m^2}  \frac{\hat \omega_0^2}{\hat \omega_s^2 - \hat \omega_{n_\ast}^2} \cos(\hat k_{n_\ast} \hat x) \cos(\hat \omega_s \hat t) \,, \\
		\hat \phi_1^{(f)}(\hat t, \hat x) &= -\frac{I}{m^2}  \frac{\hat \omega_0^2 + \hat \omega_s^2 - \hat \omega_{n_\ast}^2}{\hat \omega_s^2 - \hat \omega_{n_\ast}^2} \cos(\hat k_{n_\ast} \hat x) \cos(\hat \omega_{n_\ast} \hat t) \,.
	\end{align}
Substituting the definition of each quantity, 
the solutions can be rewritten as,
	\begin{align}
		\hat \phi_1^{(s)}(\hat t, \hat x) &= -\left[ \frac{\hat k_{\rm ad}^2}{(1-c_s^2)\hat k_{n_\ast}^2+ \hat k_{\rm ad}^2} \right] \frac{\hat \rho_1}{\hat m^2} \,,  \\
		\hat \phi_1^{(f)}(\hat t, \hat x) &= -\frac{I  \cos(\hat k_{n_\ast} \hat x)}{\hat m^2} \left[ \frac{ \hat k_{n_\ast}^2}{\hat k_{n_\ast}^2 + (1-c_s^2)^{-1}\hat k_{\rm ad}^2} \right] \cos(\hat \omega_f t) \,,
	\end{align}
where $\hat \omega_f = \hat \omega_{n_\ast} = \sqrt{k_{n_\ast}^2 + k_{\rm ad}^2}$.

\subsubsection{Static initial condition}

As we have explained in Sec.~\ref{linear_adiabatic}, when the density contrast is large, 
it is more appropriate to assume the static initial condition
		\begin{align}
		\hat \phi_1(\hat t=0, \hat x) = \hat \phi_s \,,\quad \pd_{\hat t} \hat \phi_1(\hat t=0, \hat x) = 0 \,,
	\end{align}
where $\hat \phi_s$ is a static solution for a given initial density profile. 
At the linear order, the static solution $\hat \phi_s$ is a solution of the differential equation
	\begin{align}
		\pdif{2}{\hat\phi_1}{x} - \frac{1}{\epsilon} \frac{a^2}{L^2}\left( \hat m^2\hat\phi_1 + \hat\rho_1 \right) = 0
	\end{align}
and given by
	\begin{align}
		 \hat \phi_s = -\frac{1}{\hat \omega_{n_\ast}^2} \frac{1}{\epsilon} \frac{a^2}{L^2} \hat \rho_1 = -\left(\frac{\hat \omega_0}{\hat \omega_{n_\ast}}\right)^2 \frac{\hat \rho_1}{\hat m^2} 
	\end{align}
in the Fourier space. 
Imposing the static initial condition, 
the homogeneous solution is determined as
	\begin{align}
		\hat \phi_{\rm hom}(\hat t, \hat x) = -\frac{I\cos(\hat k_{n_\ast} \hat x)}{\hat m^2}\left(\frac{\hat \omega_0}{\hat \omega_{n_\ast}}\right)^2\cos(\hat \omega_{n_\ast} \hat t) \,,
	\end{align}
and the slow and fast modes are given by	
	\begin{align}
		\hat \phi_1^{(s)}(\hat t, \hat x) &= \frac{I}{m^2}  \frac{\hat \omega_0^2}{\hat \omega_s^2 - \hat \omega_{n_\ast}^2} \cos(\hat k_{n_\ast} \hat x) \cos(\hat \omega_s \hat t) \,, \\
		\hat \phi_1^{(f)}(\hat t, \hat x) &= -\frac{I\cos(\hat k_{n_\ast} \hat x)}{\hat m^2}\left(\frac{\hat \omega_0}{\hat \omega_{n_\ast}}\right)^2
		\left[ \left( \frac{\hat \omega_s}{\hat \omega_{n_\ast}}\right)^2\frac{\cos(\hat \omega_{n_\ast} \hat t)}{1- \hat \omega_s^2/\hat \omega_{n_\ast}^2} \right] \,.
	\end{align}
Substituting the definition of each quantity, 
the solutions can be rewritten as,		
	\begin{align}
		\hat\phi_1^{(s)}(\hat{t}, \hat{x}) &= -\left[ \frac{\hat k_{\rm ad}^2}{(1-c_s^2)\hat k^2 + \hat k_{\rm ad}^2 } \right] \frac{\hat \rho_1}{\hat m^2}, \\
		\hat\phi_1^{(f)}(\hat{t}, \hat{x}) &=  -\frac{I \cos(\hat k \hat x)}{\hat{m}^2}  \left( \frac{c_s^2 \hat k^2}{k^2+k_{\rm ad}^2} \right) \left[ \frac{\hat k_{\rm ad}^2}{(1-c_s^2)\hat k^2 + \hat k_{\rm ad}^2 } \right]  \cos\left (\hat \omega_f \hat t \right). 
	\end{align}
	
\subsection{General density profile}
Having finished the analysis for the standing acoustic wave,
let us now see what determines the excitation of the fast mode for a general density profile $\hat \rho_1(\hat t, \hat x)$. 
Since we only consider an even function for the density profile $\hat \rho_1(\hat t, \hat x)$ in the analysis, 
we consider the profile decomposed as
	\begin{align}\label{eq:general profile}
		\hat \rho_1(\hat t, \hat x) = I \sum_{n \in \iz^+} c_n(\hat t) \cos(\hat k_n \hat x) \,,
	\end{align}
where the normalization $I$ is determined through
	\begin{align}
		\sum_{n \in \iz^+} |c_n(\hat t=0)|^2 = 1 \,.
	\end{align}
For a general profile (\ref{eq:general profile}), the solution (\ref{eq:green sol}) can be written as
   	 \begin{align}
		\hat{\phi}_1(\hat{t}, \hat{x}) - \hat{\phi}_{\rm hom}(\hat{t},\hat{x})
		= -\frac{\hat \omega_0^2 I}{\hat m^2} \sum_{n \in \iz} \left\{ \frac{\cos(\hat k_n \hat x)}{\hat\omega_n} \int_{0}^{\hat{t}} {\rm d}\hat{t}' \sin[\omega_n(\hat{t}-\hat{t}')] c_n(\hat t) \right\} \,.
	\end{align}
In the time integral,
	\begin{align}\label{eq:time integral}
		 \int_{0}^{\hat{t}} {\rm d}\hat{t}' \sin[\omega_n(\hat{t}-\hat{t}')] c_n(\hat t) = \frac{e^{-i \hat \omega_n \hat t}}{2i}\int_{0}^{\hat t} {\rm d}\hat t' e^{i \hat \omega_n \hat t'} c_n(\hat t') + \text{c.c} \,,
	\end{align}
the factor $e^{i\omega_n t'}$ rapidly oscillates, while the Fourier component of the density profile $c_n(t)$ slowly varies.
Then, as in the method of steepest descent, 
we assume that $c_n(t)$ is an analytic function and evaluate the integral (\ref{eq:time integral}) by deforming the integration path to $[0, 0+i\infty] \cup [0+i\infty, \hat t+i\infty] \cup [\hat t+i\infty, \hat t] $,
	\begin{align}
		\frac{e^{-i \hat \omega_n \hat t}}{2i} \int_{0}^{\hat t} {\rm d}\hat \hat t' e^{i \hat \omega_n \hat t'} c_n(\hat t')
		 = \frac{e^{-i \hat \omega_n \hat t}}{2} \int_{0}^\infty {\rm d}p~ e^{-\hat \omega_n p} c_n(ip) 
		 - \frac{1}{2}\int_{0}^\infty {\rm d}p~ e^{-\hat \omega_n p} c_n(\hat t+ip) \,,
	\end{align}
where we have dropped the contribution from the path $[0+i\infty, t+i\infty]$ assuming that $\lim_{p \to \infty} e^{-\omega_n p} c_n(\hat{t}'+ip) =0$ for $0 < \hat t' < \hat t$.
In this expression, 
the first and second terms depend on time through the factor $e^{-i \hat \omega_n \hat t}$ and $c_n(\hat t+ip)$, respectively. 
Therefore, 
it gives the decomposition into the fast and slow modes. 
In terms of the real and imaginary parts of $c_n(ip)$
	\begin{align}
		c_n(ip) = a_n(p) + i b_n(p) \,,
	\end{align}
the contribution from the first term can be written as
	\begin{align}
		&\frac{e^{-i \hat \omega_n \hat t}}{2} \int_{0}^\infty {\rm d}p~ e^{-\hat \omega_n p} c_n(ip) + \text{c.c} = \nonumber \\
		&\qquad \qquad \cos(\hat\omega_n t) \left(\int_{0}^\infty {\rm d}p e^{-\hat\omega_n p} a_n(p) \right) + \sin(\hat\omega_n t) \left(\int_{0}^\infty {\rm d}p e^{-\hat\omega_n p} b_n(p) \right) \,.
	\end{align}
	
The static solution for the density profile (\ref{eq:general profile}) is given by
	\begin{align}
		\hat \phi_s(\hat x) = -\sum_{n \in \iz^+} \left(\frac{\hat \omega_0}{\hat \omega_n}\right)^2 \frac{c_n(0)\cos(\hat k_n \hat x)}{\hat m^2} \,.
	\end{align}
Imposing the static initial condition (\ref{static ini linear}), 
the fast mode is finally given by
	\begin{align}
		\hat \phi_1^{(f)}(\hat t, \hat x) = 
		\frac{I}{\hat m^2}\sum_{n \in \iz^+} \cos(\hat k_n \hat x)\left(\frac{\hat \omega_0}{\hat \omega_n}\right)^2\left[ A \cos(\hat \omega_n \hat t) + B \sin(\hat \omega_n \hat t)\right] \,,
	\end{align}
where
	\begin{align}
		A &\equiv  -c_n(0) + \int_{0}^\infty {\rm d}\tau e^{-\tau} a_n\left(\frac{\tau}{\hat \omega_n}\right) \,, \\
		B &\equiv  \int_{0}^\infty {\rm d}\tau e^{-\tau} b_n\left(\frac{\tau}{\hat \omega_n}\right) \,,
	\end{align}
with $\tau \equiv \omega_n p$. 
Making a Taylor series expansion of $c_n(t)$,  we can write the amplitudes $A$ and $B$ also as
	\begin{align}
		A &=  \frac{1}{\hat\omega_n^2}\sum_{N =0}^\infty \frac{c_n^{(2N+2)}(0)}{\hat\omega_n^{2N}} \simeq \frac{\ddot{c}_n(0)}{\hat \omega_n^2} \,, \\
		B &=  \frac{1}{\hat\omega_n}\sum_{N =0}^{\infty} \frac{c_n^{(2N+1)}(0)}{\hat\omega_n^{2N}} \simeq  \frac{\dot{c}_n(0)}{\hat \omega_n} \,.
	\end{align}
In the second equalities, we have neglected the higher-order terms because they are suppressed by $\hat \omega/\hat \omega_n \ll 1$ 
for a typical variation scale of the density perturbation $\hat \omega$.
When we impose the initial condition (\ref{nonlinear_ini}), 
the first derivative $\dot{c}_n(0)$ vanishes and the first mode can be approximated as
	\begin{align}
		\hat \phi_1^{(f)}(t, x) &\simeq 
		\frac{I}{\hat m^2}\sum_{n \in \iz^+} \cos(\hat k_n \hat x)\left(\frac{\hat \omega_0}{\hat \omega_n}\right)^2 \left[ \frac{\ddot{c}_n(0)}{\hat \omega_n^2} \right] \cos(\hat \omega_n \hat t) \nonumber \\
\label{general_f}
		&=\frac{I}{\hat m^2}\sum_{n \in \iz^+} \cos(\hat k_n \hat x)\left(\frac{\hat{k}_{\rm ad}^2}{\hat{k}_n^2+\hat{k}_{\rm ad}^2}\right)^2 \left[ \frac{\ddot{c}_n(0)}{\hat{k}_n^2+\hat{k}_{\rm ad}^2} \right] \cos(\hat \omega_n \hat t) \,.
	\end{align}

\section{Static profile}
\label{st_profile}

\begin{figure}[t]
\centering
		\includegraphics[width=8cm]{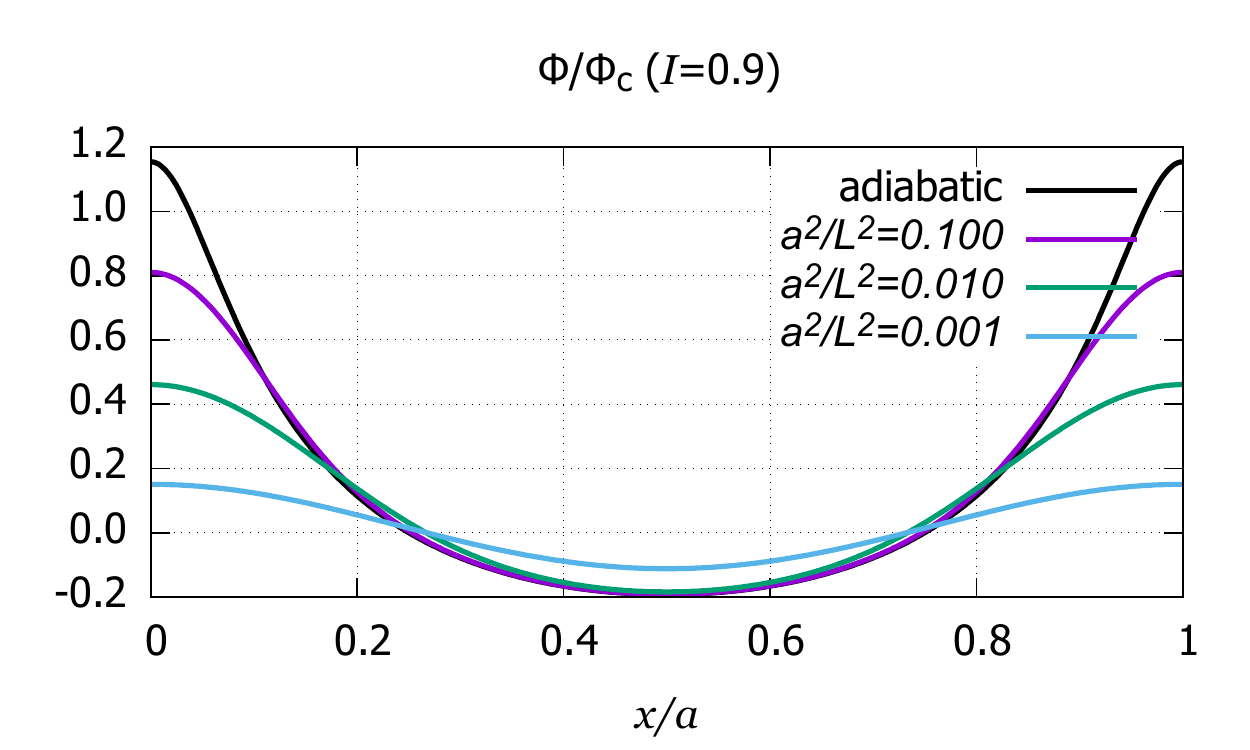}~
		\includegraphics[width=8cm]{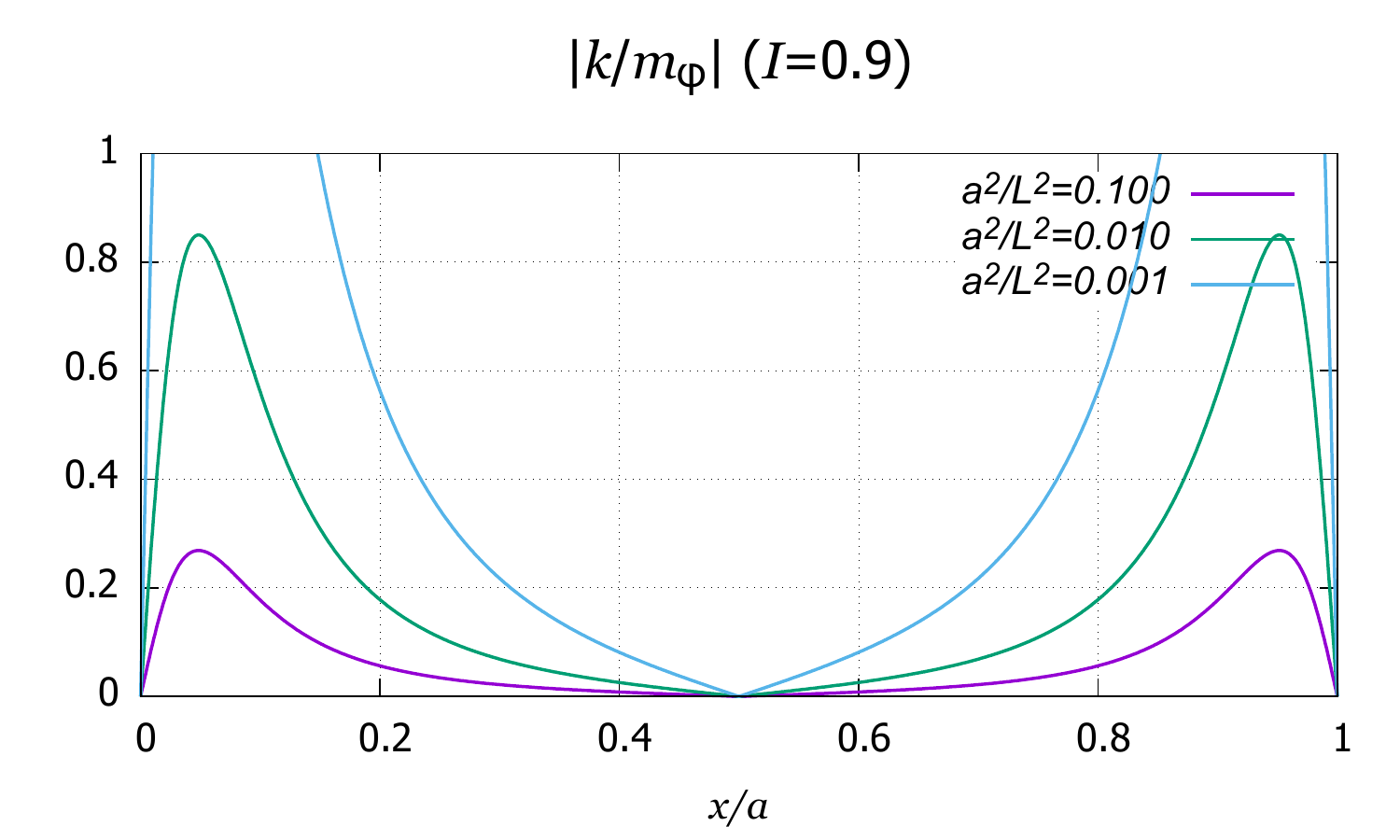}
\centering
\caption{
\baselineskip5mm
Left panel: static profiles for several values of $a^2/L^2$ compared to the adiabatic profile.
Right panel:
the degree of violation of the adiabatic approximation (see Eq.~\eqref{adiabatic_cond}), where we define $k$ and $m_{\phi}$ as $\hat k=\partial_{\hat x}\hat\phi_1^{(\rm ad)}/\hat\phi_1^{(\rm ad)}=\partial_{\hat x}\hat\rho_1/((n+1)(1+\hat\rho_1))$ and $\hat m_{\phi}=\sqrt{(n+1)a^2/L^2\cdot\epsilon\cdot(1+\hat\rho_1)^{(n+2)/(n+1)}}$.
We set $\epsilon=10^{-4}, c_s^2/\gamma=0.1$.}
\label{initial_profile}
\end{figure}

The adiabatic mode of the scalar field diverges if there is a vacuum region.
Instead,
in the numerical calculations, 
we take the static solution of $\phi$ at each moment as the slow mode, which does not diverge even in vacuum region as we saw in Sec.~\ref{numerical}.
To examine
the difference of the static profile from the adiabatic profile, we show the static profile for several initial density amplitudes in Fig.~\ref{initial_profile}.
We can observe that the static profile deviates from the adiabatic one, and the deviation becomes larger in regions where the adiabatic condition~(\ref{adiabatic_cond}) is violated.
As the initial amplitude approaches $1$, the static profile converges to some profile with a finite amplitude.
We can obtain the amplitude for an extremely inhomogeneous density profile such as aligned infinitely thin shells~\cite{Nakamura2019}, 
and expect it approximately gives an upper bound for the amplitude. 
For $n=2$, we gave the following maximum amplitude~\cite{Nakamura2019}: 
\b
\hat\phi_{\rm max}\sim \left(\frac{1}{4\epsilon}\frac{a^2}{L^2}\right)^{1/4},
\label{phi_max}
\e
which will be used to estimate the maximum amplitude of the fifth force due to the fast mode in the next section.

\section{Initial condition with vanishing scalar field}
\label{other_initial}
We may consider a situation in which the density distribution suddenly changes within very short time scale 
so that the scalar field configuration cannot follow it. 
With such a case in mind, 
let us consider the following initial condition:
\b
\hat\rho_1|_{t=0}=-I\cos(2\pi x), 
\qquad
\hat\phi_1=0,
\qquad
v|_{ t=0}=\partial_t \phi|_{t=0}=0.
\e
\begin{figure}[t]
\centering
		\includegraphics[width=16cm]{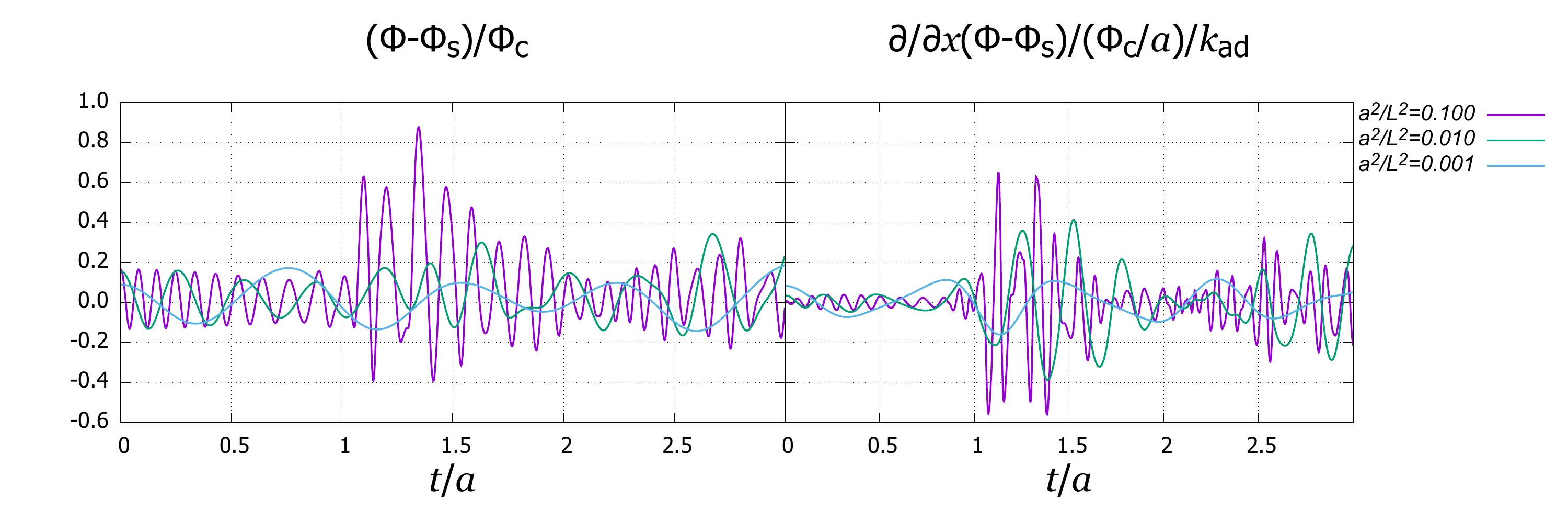}
		\caption{
\baselineskip5mm
The fast mode amplitude and the fifth force value for $I=0.9$ with $a^2/L^2= 0.1, 0.01, 0.001$.
We set $\epsilon=10^{-4}$, $c_s^2/\gamma=0.1$.
}
\label{evolution_phi_zero}
\centering
\end{figure}
From Fig.~\ref{initial_profile} and the left panel of  Fig.~\ref{evolution_phi_zero}, 
we can see that the maximum amplitude of the fast mode (left panel of Fig.~\ref{evolution_phi_zero}) is 
roughly given by the maximum value of the static profile corresponding to the initial density profile. 
The right panel show the spatial derivative of the fast mode divided by $k_{\rm ad}$.
It shows that we can estimate the derivative of the fast mode by multiplying $k_{\rm ad}$, 
similarly to the estimate made in Sec.~\ref{sec:comparison}. 

Let us estimate
an upper bound on the the fifth force. 
Combining Eq.~\eqref{phi_max} with the estimation of the spatial derivative, we obtain
\begin{align}
F_{\phi{\rm max}}&\sim\frac{1}{\sqrt{2}}\left(\frac{a}{L}\right)^{3/2}\epsilon^{1/4}~.
\label{fifth_max}
\end{align}

\end{appendix}

\bibliographystyle{apsrev4-1}

\end{document}